\documentclass{emulateapj}
\usepackage{graphicx}

\usepackage{ulem}
\def\kms{\ifmmode{\rm km\thinspace s^{-1}}\else km\thinspace s$^{-1}$\fi}
\def\Teff{\ifmmode{T_{\rm eff}}\else $T_{\rm eff}$\fi}

\shorttitle{IM Vir}
\shortauthors{Morales et al.}

\begin{document}

\title{Absolute dimensions of the G7+K7 eclipsing binary star
IM~Virginis: Discrepancies with stellar evolution models}

\author{Juan Carlos Morales\altaffilmark{1,$\star$},
Guillermo Torres\altaffilmark{2}, Laurence A. Marschall\altaffilmark{3} 
and William Brehm\altaffilmark{3}}


\altaffiltext{1}{Institut d'Estudis Espacials de Catalunya (IEEC), Edif.
Nexus, C/ Gran Capit\`a, 2-4, 08034 Barcelona, Spain; morales@ieec.uab.es}
\altaffiltext{2}{Harvard-Smithsonian Center for Astrophysics, 60 Garden Street,
Cambridge, MA 02138, USA}
\altaffiltext{3}{Department of Physics, Gettysburg College,
300 North Washington Street, Gettysburg, PA 17325, USA}
\altaffiltext{$\star$}{Visiting fellowship at Harvard-Smithsonian Center for
Astrophysics}

\begin{abstract}

We report extensive spectroscopic and differential photometric $BV\!RI$
observations of the active, detached, 1.309-day
double-lined eclipsing binary IM~Vir, composed of a G7-type primary
and a K7 secondary. With these observations we derive accurate
absolute masses and radii of $M_1 = 0.981 \pm 0.012~M_{\sun}$, $M_2 =
0.6644 \pm 0.0048~M_{\sun}$, $R_1 = 1.061 \pm 0.016~R_{\sun}$, and
$R_2 = 0.681 \pm 0.013~R_{\sun}$ for the primary and secondary, with
relative errors under 2\%. The effective temperatures are $5570 \pm
100$~K and $4250 \pm 130$~K. The significant difference in mass makes
this a favorable case for comparison with stellar evolution theory. We
find that both stars are larger than the models predict, by 3.7\% for
the primary and 7.5\% for the secondary, as well as cooler than
expected, by 100~K and 150~K, respectively. These discrepancies are in
line with previously reported differences in low-mass stars, and are
believed to be caused by chromospheric activity, which is not
accounted for in current models. The effect is not confined to
low-mass stars: the rapidly-rotating primary of IM~Vir joins the
growing list of objects of near-solar mass (but still with convective
envelopes) that show similar anomalies. The comparison with the models
suggests an age of 2.4~Gyr for the system, and a metallicity of
[Fe/H] $\approx -0.3$ that is consistent with other indications, but
requires confirmation.

\end{abstract}


\section{Introduction}
\label{sec:introduction}

Our knowledge of stellar structure and evolution rests heavily on the
comparison between theory and observation. Double-lined eclipsing
binaries (hereafter EBs) have long been at the center of this process,
since they allow the mass -- the most fundamental of all stellar
properties -- as well as the radius to be determined to very high
precision (and accuracy), often as good as 1\%, independently of the
distance and independently of any calibrations. Such high precision
enables stringent tests of theory, as described, e.g., by
\cite{Andersen1991}.  In the last decade or so it has become clear
that stars in the lower main-sequence show significant discrepancies
when compared to standard models. Studies of several key systems have
shown unambiguously that the radii predicted by the models are
systematically up to $\sim$10\% too small, and the temperatures
$\sim$5\% too high \citep[see][and references therein]{Ribas2008}. The
deviations are commonly attributed to the high level of chromospheric
activity present in these systems. Orbital periods are typically
short, and as a result tidal forces tend to synchronize the
components' rotation with the orbital motion. The rapid rotation, in
turn, induces the activity, which can manifest itself in the form of
copious X-ray emission, flaring, H$\alpha$ and Ca~II H and K emission,
spottedness, and other effects.

According to the recent compilation by \cite{Torresetal:09}, among the
eclipsing binaries with at least one component below 0.8~M$_{\odot}$
only five have reliable mass and radius determinations with relative
errors that are smaller than 3\%. These are, in order of decreasing
mass, UV~Psc \citep{Popper:97}, GU~Boo \citep{LopezMorales2005},
YY~Gem \citep{Torres2002}, CU~Cnc \citep{Ribas2003}, and CM~Dra
\citep{Morales2009}. A number of other similar systems are known, but
are not yet at the same level of precision \emph{and} accuracy. For
further progress it is therefore important to either improve the
determinations in the latter cases, or to find new ones where those
conditions are met. The eclipsing system reported here is one of
these, which in addition presents the largest difference in mass
between the components, providing in principle extra leverage for the
comparison with models.

IM~Virginis (also HD~111487, 1E~1247.0$-$0548, $\alpha = 12^{\rm
h}49^{\rm m}38.\!^{\rm s}70$, $\delta = -6^{\circ}04'44.\!''9$,
J2000.0; \ion{G7}{5}, $V = 9.57$) was detected as an X-ray source with
the \textit{Einstein} Observatory by \cite{Helfand1982}. The
radial-velocity variability was found by \cite{Silva1987}, and
subsequent spectroscopic and photometric studies carried out by
\cite{Marschall1988, Marschall1989} confirmed IM~Vir to be both
double-lined and eclipsing, and to be composed of a G7 star and a
late-K or early-M star, thus piquing our interest. The orbital period
was estimated as 1.3085 days.

Very little data on this binary have been published since, other than
sparse photometry and occasional reports on the chromospheric activity
and X-ray flaring \citep{Strassmeier1993, Pandey2008}. We present here
extensive spectroscopic and differential photometric measurements that
allow us to determine its fundamental properties with very high
accuracy. The observations are presented in
\S\thinspace\ref{sec:spectroscopy} and \S\thinspace\ref{sec:lc_obs},
followed by a detailed light-curve analysis in
\S\thinspace\ref{sec:lc_analysis} accounting for the presence of
spots. The absolute dimensions are derived in
\S\thinspace\ref{sec:absolute} with careful consideration of potential
sources of systematic error, essential for the results to be
useful. The comparison with models of stellar evolution is discussed
in \S\thinspace\ref{sec:discussion}, and we summarize our conclusions
in \S\thinspace\ref{sec:conclusions}.

\section{Spectroscopic observations and reductions}
\label{sec:spectroscopy}

IM~Vir was observed at the Harvard-Smithsonian Center for Astrophysics
(CfA) with three nearly identical echelle spectrographs on the 1.5-m
Tillinghast reflector at the F.\ L.\ Whipple Observatory (Mount
Hopkins, Arizona), the 1.5-m Wyeth reflector at the Oak Ridge
Observatory (Harvard, Massachusetts), and the 4.5-m equivalent
Multiple Mirror Telescope (also on Mount Hopkins, Arizona) prior to
its conversion to a 6.5-m monolithic telescope.  Photon-counting
intensified Reticon detectors \citep[`Digital
Speedometers';][]{Latham:85, Latham:92} were used in each case to
record a single 45\,\AA\ echelle order centered at a wavelength of
5188.5\,\AA, featuring the gravity-sensitive lines of the
\ion{Mg}{1}~b triplet. The resolving power provided by this setup is
$\lambda/\Delta\lambda\approx 35,\!000$. Nominal signal-to-noise (S/N)
ratios for the 138 spectra we obtained range from 13 to 58 per
resolution element of 8.5~\kms. The first observation was taken in
1984 January 1, and monitoring continued until 2009 May 10. A handful
of the early spectra are the same ones included in the work of
\cite{Silva1987}, who discovered the radial-velocity variability, but
have been re-reduced and analyzed here with much improved methods, as
we now describe.

All of our spectra are double-lined, but the secondary is
comparatively quite faint, accounting for only 6\% of the light of the
primary (see below).  Radial velocities for both stars were derived
using TODCOR, a two-dimensional cross-correlation technique introduced
by \cite{Zucker:94}. This method uses two templates, one for each
component of the binary, which we selected from a large library of
synthetic spectra based on model atmospheres by R.\ L.\ Kurucz
\citep[see][]{Latham:02}. These templates have been calculated for a
wide range of effective temperatures ($T_{\rm eff}$), surface
gravities ($\log g$), rotational velocities ($v \sin i$ when seen in
projection), and metallicities ([m/H]). Following \cite{Torres:02},
the optimum templates were selected by running extensive grids of
cross-correlations with TODCOR, seeking to maximize the average
correlation weighted by the strength of each exposure.  Because of the
faintness of the secondary, only the parameters for the bright star
can be determined reliably from our spectra. Initially we assumed
$\log g = 4.5$, and determined $T_{\rm eff}$ and $v \sin i$ for a
range of metallicities between [m/H] $= -1.0$ and [m/H] $= +0.5$. The
optimal values were found by interpolation in [m/H]. We then repeated
the correlations for a lower value of $\log g = 4.0$ in order to
bracket the estimate from the analysis described later ($\log g =
4.379$). Interpolation in $\log g$ to this final value resulted in
$T_{\rm eff} = 5570 \pm 100$~K and $v \sin i = 43 \pm 2$~\kms. The
formal metallicity that maximizes the correlation is [m/H] = $-0.1$.
While this suggests an overall abundance close to solar, the
uncertainty is likely to be at least 0.25~dex, and we consider the
estimate primarily as a free parameter included to optimize the match
between the synthetic templates and the observed spectra.  We note,
furthermore, that because of our narrow spectral window, there is a
strong correlation between temperature and metallicity. In this case,
however, the $T_{\rm eff}$ value reported above is consistent with
various photometric estimates described below in
\S\thinspace\ref{sec:absolute}, in turn lending more credence to the
metallicity estimate. For the final velocity determinations we adopted
the set of template parameters closest to the values above that
maximizes the average correlation ($T_{\rm eff} = 5750$~K, $v \sin i =
40$~\kms, $\log g = 4.5$, and solar composition).  Small differences
in these template parameters compared to the interpolated values do
not affect the velocities significantly. For the secondary template we
adopted the same metallicity, and parameters consistent with the
results from \S\thinspace\ref{sec:absolute}: $T_{\rm eff} = 4250$~K,
$v \sin i = 25$~\kms, $\log g = 4.5$.

In addition to the velocities, we have determined the light ratio
between the components at the mean wavelength of our observations,
following \cite{Zucker:94}, accounting for the difference in line
blocking between the primary and the much cooler secondary.\footnote{This
is to correct the ratio of continuum heights for the fact that the lines of
the secondary are intrinsically stronger, and subtract proportionately
more flux from the spectral window.} The value we obtain is
$\ell_2/\ell_1 = 0.06 \pm 0.01$.

\begin{figure}[t]
\centering
\includegraphics[width=0.8\columnwidth]{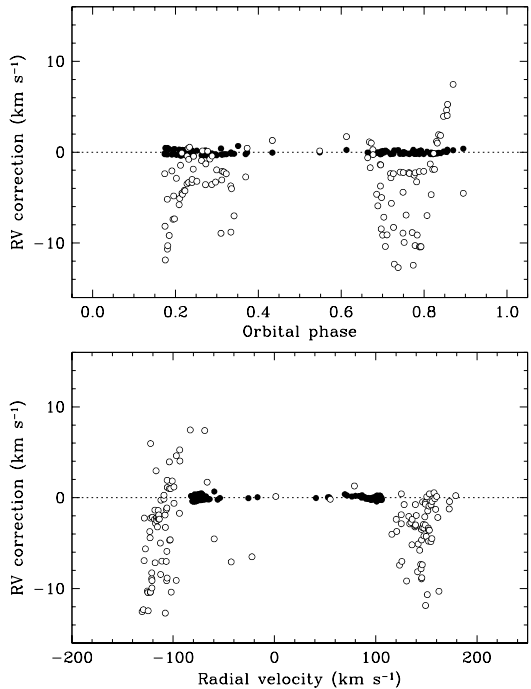}
\caption{Systematic corrections applied to the raw radial velocity
measurements as a function of orbital phase (top) and radial velocity
(bottom).  Filled symbols represent the primary, and open symbols are
for the secondary.}
\label{fig:cortodcor}
\end{figure}

Although TODCOR significantly reduces systematic errors in the radial
velocities caused by line blending, residual effects can remain as a
result of shifts of the spectral lines in and out of our narrow
spectral window as a function of orbital phase.  Experience has shown
that these effects must be examined on a case-by-case basis. We
investigated them here by means of numerical simulations similar to
those described by \cite{Latham:96} \citep[see also][]{Torres:97,
Torres:00}. We generated synthetic composite spectra matching our
observations by combining copies of the primary and secondary
templates used above, shifted to the appropriate velocities for each
of the exposures as predicted by a preliminary orbital solution, and
scaled to the observed light ratio. These synthetic observations were
then processed with TODCOR in exactly the same way as the real
spectra, and the resulting velocities were compared with the input
shifts.  The differences for IM~Vir are shown in
Figure~\ref{fig:cortodcor}, and are less than 0.5~\kms\ for the
primary but reach values as large as 13~\kms\ for the secondary.
Similarly large differences have been found occasionally for other
systems using the same instrumentation \citep[e.g., AD~Boo, GX~Gem,
VZ~Cep;][]{Clausen:08, Lacy:08, Torres:09}. We have applied these
differences as corrections to the raw velocities of IM~Vir. The impact
on the masses, however, is very small: only 0.26\% for the primary and
0.14\% for the secondary. The reason for this is that the large
corrections for the secondary are similar and of the same sign
at both quadratures, therefore
amounting mostly to an overall systematic shift rather than a change in
the velocity semi-amplitude. Similar adjustments based on the same
simulations were made to the light ratio, and are already included in
the value reported above.

\begin{deluxetable*}{lccccccc}
\tablewidth{0pc}
\tablecaption{Radial velocity measurements of IM~Vir.\label{tab:rvs}}
\tablehead{\colhead{HJD} &
\colhead{$RV_1$} &
\colhead{$RV_2$} &
\colhead{$\sigma_1$} &
\colhead{$\sigma_2$} &
\colhead{$(O\!-\!C)_1$} &
\colhead{$(O\!-\!C)_2$} &
\colhead{} \\
\colhead{\hbox{~~(2,400,000$+$)~~}} &
\colhead{(\kms)} &
\colhead{(\kms)} &
\colhead{(\kms)} &
\colhead{(\kms)} &
\colhead{(\kms)} &
\colhead{(\kms)} &
\colhead{Phase}}
\startdata
 45719.9949\dotfill  &  \phn$-$67.58 &  $+$144.06  &    1.54   &   13.01  &   $+$0.30 &  \phn$+$9.61  &   0.1661 \\
 45743.0126\dotfill  &     $+$105.93 &  $-$116.81  &    0.83   &\phn6.98  &   $+$1.08 &  \phn$+$3.76  &   0.7555 \\
 45750.8945\dotfill  &     $+$100.91 &  $-$122.28  &    1.58   &   13.30  &   $-$2.51 &  \phn$-$3.82  &   0.7785 \\
 45754.0581\dotfill  &  \phn$-$74.09 &  $+$148.80  &    1.54   &   12.97  &   $+$1.10 &  \phn$+$3.56  &   0.1961 \\
 45754.0603\dotfill  &  \phn$-$71.28 &  $+$149.09  &    1.90   &   16.05  &   $+$4.23 &  \phn$+$3.38  &   0.1977 \\
\enddata
\tablecomments{This table is available in its entirety in a machine-readable form in the online journal. A portion is shown here for guidance regarding its form and content.}
\end{deluxetable*}

The stability of the zero-point of the CfA velocity system was
monitored by means of exposures of the dusk and dawn sky, and small
run-to-run corrections were applied in the manner described by
\citet{Latham:92}. The final velocities, including these offsets as
well as the corrections for systematics, are listed in
Table~\ref{tab:rvs} along with their uncertainties. The median values
for the velocity errors are 0.90~\kms\ for the primary and 7.6~\kms\
for the secondary.

Preliminary single-lined orbital solutions carried out separately for
the primary and secondary indicated a significant zero-point
difference (i.e., a difference in the systemic velocity $\gamma$),
which is often seen by many investigators in cases where there is a
slight mismatch between the spectra of the real stars and the
templates used for the cross-correlations \citep[see,
e.g.,][]{Popper:00, Griffin:00}. Numerous experiments with other
templates did not remove the offset, which is about 4~\kms\ (the
secondary velocities being systematically larger). In IM~Vir this
most likely arises because of the cool temperature of the secondary,
and the fact that synthetic templates for such stars become
increasingly unrealistic due to missing opacity sources in the
models. In order to prevent this offset from affecting the velocity
semiamplitudes, we have therefore accounted for it by including it as
an additional free parameter in our double-lined solution. Effectively
this means we allow each component to have its own center-of-mass
velocity, without significantly affecting the velocity semiamplitude of
the component, and hence without affecting the individual mass
determination. On the other hand, the offset implies additional
uncertainty in the true center-of-mass velocity of the binary, beyond the
formal errors listed below. We do not expect a significant template
mismatch for the solar-type primary, so the systemic velocity is most
likely to be closer to the value for that star.

\begin{figure}[t]
\centering
\includegraphics[width=0.9\columnwidth]{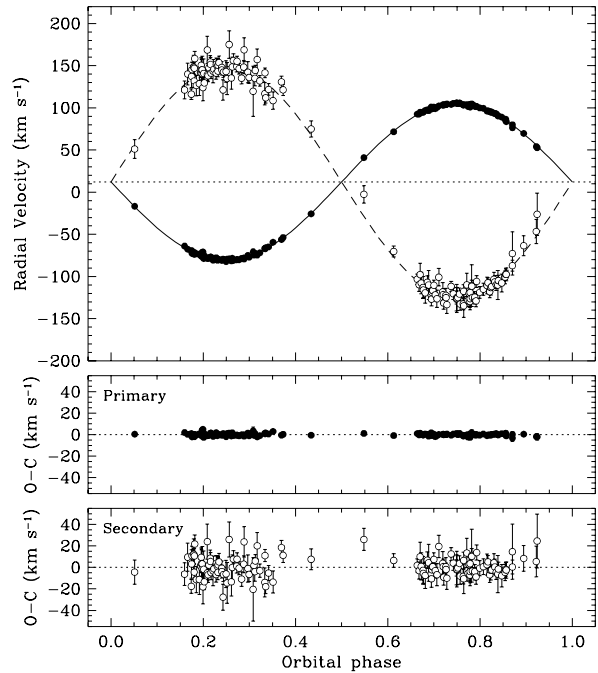}
\caption{Phase-folded radial velocity observations for IM~Vir (filled
circles for the primary, open circles for the secondary), along with
our best-fit model. Phase 0.0 corresponds to primary eclipse. The
residuals from the fit are shown in the bottom panels.}
\label{fig:specorbit}
\end{figure}

The results are presented in Table~\ref{tab:specorbit}, in which the
measurements have been weighted according to their uncertainties in
the usual way, and the errors rescaled by iterations to achieve a
reduced $\chi^2$ value near unity, separately for the primary and
secondary. No indication of significant eccentricity was found, as expected
for such a short orbital period, and the final orbit was therefore
considered circular. The observations and computed curve are displayed
graphically in Figure~\ref{fig:specorbit}, along with the residuals,
which are also listed in Table~\ref{tab:rvs}.

\begin{deluxetable}{lc}
\tablewidth{0pc}
\tablecaption{Spectroscopic orbital solution for IM~Vir.\label{tab:specorbit}}
\tablehead{\colhead{~~~~~~~~~~~~~Parameter~~~~~~~~~~~~~~} & \colhead{Value}}
\startdata
\multicolumn{2}{l}{Adjusted quantities\hfil} \\
~~~~$P$ (days)\dotfill                           & 1.30861497~$\pm$~0.00000034  \\
~~~~$T_{\rm I}$ (HJD$-2,\!400,\!000$)\tablenotemark{a}\dotfill    & 52402.87420~$\pm$~0.00052\phm{2222}     \\
~~~~$K_{1}$ (\kms)\dotfill                       & 92.684~$\pm$~0.088\phn    \\
~~~~$K_{2}$ (\kms)\dotfill                       & 136.84~$\pm$~0.74\phn\phn    \\
~~~~$e$\tablenotemark{b}\dotfill                 & 0 \\
~~~~$\gamma$ (\kms)\tablenotemark{c}\dotfill     & +12.221~$\pm$~0.082\phn\phs \\
~~~~$\Delta RV$ (\kms)\tablenotemark{d}\dotfill  & $-$3.96~$\pm$~0.69\phs \\
\multicolumn{2}{l}{Derived quantities\hfil} \\
~~~~$M_1\sin^3 i$ ($M_{\sun}$)\dotfill           &  0.978~$\pm$~0.012            \\
~~~~$M_2\sin^3 i$ ($M_{\sun}$)\dotfill           &  0.6621~$\pm$~0.0044          \\
~~~~$q\equiv M_{2}/M_{1}$\dotfill                & 0.6773~$\pm$~0.0037 \\
~~~~$a_1\sin i$ (10$^6$ km)\dotfill            & 1.6678~$\pm$~0.0016 \\
~~~~$a_2\sin i$ (10$^6$ km)\dotfill            & 2.462~$\pm$~0.013 \\
~~~~$a\sin i$ (10$^6$ km)\dotfill              & 4.130~$\pm$~0.013 \\
\multicolumn{2}{l}{Other quantities pertaining to the fit\hfil} \\
~~~~$\sigma_{1}$ (\kms)\dotfill                  & 0.99\\
~~~~$\sigma_{2}$ (\kms)\dotfill                  & 8.39\\
~~~~$N_{\rm obs}$\dotfill                        & 138 \\
~~~~Time span (days)\dotfill                     &  9241.7
\enddata
\tablenotetext{a}{Time of primary eclipse.}
\tablenotetext{b}{Fixed; see text.}
\tablenotetext{c}{True uncertainty may be larger due to systematic
effects; see text.}
\tablenotetext{d}{Zero-point difference between the primary and
secondary velocities, in the sense primary \emph{minus} secondary.}
\end{deluxetable}

\section{Photometric observations}
\label{sec:lc_obs}

Differential photometric observations of IM~Vir were conducted in 2006
April and May using the 0.4-m Ealing Cassegrain reflector of the
Gettysburg College Observatory (GCO, Gettysburg, Pennsylvania).  The camera
was a Photometrics (Roper Scientific) CH-350 thermoelectrically-cooled
unit equipped with a SITe 003B back-illuminated scientific grade CCD
chip and Bessell $BV\!RI$ filters.  Exposure times were 120, 60, 25,
and 20 seconds, respectively. We obtained 597, 672, 630, and 631
observations in $B$, $V$, $R$, and $I$, with a cadence of about 6
minutes. The field of view of the GCO camera is approximately 18
arcmin, and since IM~Vir is a relatively bright target, this severely
limited our choice of reference stars of comparable magnitude and
color to two: BD$-$05~3573 (`comp', $\alpha = 12^{\rm h}48^{\rm
m}36.\!^{\rm s}40$, $\delta = -5^{\circ}53'33.\!''6$, J2000.0; $V =
10.39$, $B-V = 0.67$) and HD~111427 (`check', $\alpha = 12^{\rm
h}49^{\rm m}14.\!^{\rm s}94$, $\delta = -5^{\circ}49'20.\!''7$,
J2000.0; $V = 9.40$, $B-V = 0.67$). The colors of these two stars are
in fact nearly identical to that of the variable, which is $B-V =
0.66$ (see \S\thinspace\ref{sec:absolute}).

Differential photometry was performed on IM~Vir and the two reference
stars in all our images by means of MIRA-AP software
(http://www.mirametrics.com/).  We employed standard aperture
photometry techniques to derive instrumental magnitudes, setting the
radius of the measuring apertures for each night using a standard
value of 2.5 times the FWHM of the seeing disk, based on previous
curve-of-growth calibrations using the same equipment. Typical errors
as represented by the scatter of the comp$-$check differences are
0.0132~mag in $B$, 0.0124~mag in $V$, 0.0135~mag in $R$, and
0.0150~mag in $I$. The primary eclipse is $\sim$0.62~mag deep in $V$,
while the depth of the secondary is only $\sim$0.08~mag.

\begin{deluxetable}{lccc}
\tablewidth{0pc}
\tablecaption{Differential $B$-band photometry of IM~Vir.\label{tab:lcB}}
\tablehead{\colhead{HJD} &
\colhead{} &
\colhead{Comp-Var} &
\colhead{Comp-Check} \\
\colhead{\hbox{~~(2,400,000$+$)~~}} &
\colhead{Phase} &
\colhead{(mag)} &
\colhead{(mag)}} 
\startdata
53835.58358\dotfill  & 0.8288 & 0.811 & 0.966  \\
53835.58732\dotfill  & 0.8317 & 0.849 & 0.995  \\
53835.59108\dotfill  & 0.8345 & 0.852 & 1.000  \\
53835.59482\dotfill  & 0.8374 & 0.853 & 1.007  \\
53835.59857\dotfill  & 0.8403 & 0.854 & 1.005
\enddata
\tablecomments{This table is available in its entirety in a machine-readable form in the online journal. A portion is shown here for guidance regarding its form and content.}
\end{deluxetable}

\begin{deluxetable}{lccc}
\tablewidth{0pc}
\tablecaption{Differential $V$-band photometry of IM~Vir.\label{tab:lcV}}
\tablehead{\colhead{HJD} &
\colhead{} &
\colhead{Comp-Var} &
\colhead{Comp-Check} \\
\colhead{\hbox{~~(2,400,000$+$)~~}} &
\colhead{Phase} &
\colhead{(mag)} &
\colhead{(mag)}} 
\startdata
53835.58481\dotfill & 0.8298 & 0.848 & 0.994  \\
53835.58855\dotfill & 0.8326 & 0.846 & 0.999  \\
53835.59230\dotfill & 0.8355 & 0.839 & 0.989  \\
53835.59605\dotfill & 0.8383 & 0.838 & 0.998  \\
53835.60705\dotfill & 0.8468 & 0.826 & 0.985 
\enddata
\tablecomments{This table is available in its entirety in a machine-readable form in the online journal. A portion is shown here for guidance regarding its form and content.}
\end{deluxetable}

\begin{deluxetable}{lccc}
\tablewidth{0pc}
\tablecaption{Differential $R$-band photometry of IM~Vir.\label{tab:lcR}}
\tablehead{\colhead{HJD} &
\colhead{} &
\colhead{Comp-Var} &
\colhead{Comp-Check} \\
\colhead{\hbox{~~(2,400,000$+$)~~}} &
\colhead{Phase} &
\colhead{(mag)} &
\colhead{(mag)}} 
\startdata
53835.62654\dotfill & 0.8616 & 0.584 & 1.027  \\
53835.63031\dotfill & 0.8645 & 0.660 & 1.011  \\
53835.63406\dotfill & 0.8674 & 0.685 & 1.002  \\
53835.63781\dotfill & 0.8703 & 0.704 & 1.003  \\
53835.64157\dotfill & 0.8731 & 0.708 & 1.005
\enddata
\tablecomments{This table is available in its entirety in a machine-readable form in the online journal. A portion is shown here for guidance regarding its form and content.}
\end{deluxetable}

\begin{deluxetable}{lccc}
\tablewidth{0pc}
\tablecaption{Differential $I$-band photometry of IM~Vir.\label{tab:lcI}}
\tablehead{\colhead{HJD} &
\colhead{} &
\colhead{Comp-Var} &
\colhead{Comp-Check} \\
\colhead{\hbox{~~(2,400,000$+$)~~}} &
\colhead{Phase} &
\colhead{(mag)} &
\colhead{(mag)}} 
\startdata
53835.58592\dotfill & 0.8306 & 0.933 & 1.013  \\
53835.58967\dotfill & 0.8335 & 0.933 & 1.014  \\
53835.59342\dotfill & 0.8363 & 0.939 & 1.034  \\
53835.59717\dotfill & 0.8392 & 0.929 & 1.025  \\
53835.60091\dotfill & 0.8421 & 0.946 & 1.032 
\enddata
\tablecomments{This table is available in its entirety in a machine-readable form in the online journal. A portion is shown here for guidance regarding its form and content.}
\end{deluxetable}

Examination of the raw photometry revealed slight trends in the
comp$-$check differences that changed from night to night and are most
likely of instrumental origin. Neither star is known to be variable,
and we find no periodicities or other discernible patterns.
Consequently, we de-trended the IM~Vir measurements by removing the
median value of the comp$-$check differences calculated over intervals
of a few hours. Even after this correction, the sparser second half of
the data presents a systematic $\sim$0.01~mag difference (fainter) in
the out-of-eclipse light level compared to the first half, and other
occasional changes possibly due to temporal evolution of surface
inhomogeneities (spots) on the surface of one or both stars. This
would indicate a timescale for evolution of these features of a few
weeks.  For these reasons we have chosen to restrict our analysis to
the first half of the observations (spanning 22 days), which provide
complete coverage of both eclipses and are cleaner overall. The
complete data set is given for all filters in
Tables~\ref{tab:lcB}--\ref{tab:lcI} in its original form, i.e.,
without the nightly corrections, and we list also the comp$-$check
differences.

Additional photometry of IM~Vir has been reported by
\cite{Manfroid1991} in the Str\"omgren $uvby$ system. These data were
obtained some twenty years earlier than our own measurements, between
1983 May and 1986 July. Unfortunately the coverage of the eclipses is
very incomplete, as shown later, so these data are not useful for
determining the geometric parameters of the system. Nevertheless, it
is possible to extract an average time of eclipse, which we present
below, as well as brightness ratios in the different bands, which will
be used in \S\thinspace\ref{sec:absolute} to deconvolve the light of
the two stars in order to obtain photometric estimates of the
temperatures and metallicity.

\subsection{Ephemeris and times of minimum}

The ephemeris adopted for the remainder of this paper is the one
calculated from the spectroscopy, 
\begin{equation}
{\rm Min~I} = 2,\!452,\!402.87420 (52) + 1.30861497 (34) E~,
\end{equation}
which, by virtue of the 25-year radial-velocity coverage, is much more
accurate than could ever be obtained from the 64-day interval of the
photometric observations. The uncertainties are indicated above in
parentheses.

Three times of eclipse from the $BV\!RI$ photometry have been measured
by fitting the light-curve model described below (including spot
parameters) to each night with sufficient coverage of a primary or
secondary minimum, simultaneously in all four passbands. The only
adjustable parameter allowed in these fits was a time shift.  A
similar procedure was followed with the \cite{Manfroid1991}
photometry, with the primary luminosity in each band added as a free
parameter and no spots considered. The resulting eclipse timings for
IM~Vir are collected in Table~\ref{tab:minima}, along with an average
time of eclipse from the spectroscopy, and the few additional times
found in the literature.

\begin{deluxetable}{lccccc}
\tablewidth{0pc}
\tablecaption{Times of eclipse for IM~Vir.\label{tab:minima}}
\tablehead{\colhead{HJD} &
\colhead{$\sigma$} &
\colhead{} &
\colhead{} &
\colhead{O$-$C} &
\colhead{} \\
\colhead{(2,400,000$+$)} &
\colhead{(days)} &
\colhead{Type\tablenotemark{a}} &
\colhead{Instr.} &
\colhead{(days)} &
\colhead{Source\tablenotemark{c}}} 
\startdata
46042.99984                  & 0.00097    & ~I &ccd &  $-$0.00561 & 1 \\
51274.8750                   &            & ~I &ccd & +0.02690 & 2 \\
51885.9610                   &            & ~I &vis&  $-$0.01029 & 2 \\
52402.87420\tablenotemark{b} & 0.00052    & ~I &spec    & \phs0.0\phn\phn\phn\phn & 4 \\
53474.62948                  & 0.00055    & ~I &ccd &  $-$0.00038 & 3 \\
53843.65995                  & 0.00010    & ~I &ccd & +0.00067 & 4 \\
53845.62427                  & 0.00052    & II &ccd & +0.00207 & 4 \\
53877.68325                  & 0.00038    & ~I &ccd &  $-$0.00002 & 4
\enddata
\tablenotetext{a}{Eclipses labeled as `I' for primary, and `II' for secondary.}
\tablenotetext{b}{Mean epoch from the radial velocities.}
\tablenotetext{c}{Source: 1. Mean epoch from the  \cite{Manfroid1991} photometry;
2. B.R.N.O.\ database (http://var.astro.cz); 3. \cite{Ogloza2008}; 4. This work.}
\end{deluxetable}

\section{Light curve analysis} 
\label{sec:lc_analysis}

The differential photometry described in \S\thinspace\ref{sec:lc_obs}
was phase-folded with the adopted ephemeris, and analyzed with the
2003 version of the WD code \citep{WD1971, wilson1979}. For the
reasons indicated earlier, only the first half of the data were used
(22 nights worth).  The main light curve parameters adjusted are the
orbital inclination ($i$), the pseudo-potentials ($\Omega_1$ and
$\Omega_2$), the secondary effective temperature ($T_{\rm eff,2}$),
the primary luminosity ($L_1$), and a phase shift. The latter would
normally be unnecessary given the accuracy of the spectroscopic ephemeris
and the short interval of the photometric observations. However, as we
describe below, the light curves of IM~Vir present distortions we ascribe
to spots, which can cause asymmetries in the eclipses that may appear as
phase shifts. Since these distortions are modeled here only in a simplified
way, it is prudent to include a phase shift as an additional parameter to
supplement the spot model. The primary
temperature was held fixed at the spectroscopic value of 5570~K, since
the light curves do not provide a strong enough constraint on both
temperatures, but only on their ratio. Emergent intensities used in
the program were taken from model atmospheres described by
\cite{vanHamme:03}. Square-root limb-darkening coefficients for the
Johnson-Cousins $BV\!RI$ bands were interpolated from the tables by
\cite{Claret2000a}, and adjusted dynamically according to the current
temperatures and surface gravities of the stars at each iteration.
The reflection albedos were fixed at the value 0.5, appropriate for
stars with convective envelopes, and the gravity darkening exponents
were set to 0.34 for the primary and 0.40 for the secondary, following
\cite{Claret2000b}.  The orbit was initially considered to be
circular, and spin-orbit synchronization was assumed based on the
evidence presented in \S\thinspace\ref{sec:absolute}.  The photometric
measurements in the four passbands were fitted simultaneously.
Outliers were rejected by 3-$\sigma$ clipping based on preliminary
solutions.  Convergence in the final fit was considered to have been
achieved when the corrections to the elements were smaller than the
internal errors in three consecutive iterations.

As is common in eclipsing binaries with late-type components, the
light curves of IM~Vir show out-of-eclipse modulations that are
attributable to spots. There is also a slight difference in the light
level at the two quadratures. These variations can in principle be
modeled with the WD code in the approximation of circular spots of
uniform temperature.  In practice, however, such modeling is fraught
with difficulties and there is abundant literature discussing problems
of indeterminacy and non-uniqueness, particularly when using data of
limited quality \citep[see, e.g.,][and references therein]{Eker:96,
Eker:99}.  Nevertheless, because we are interested in recovering the
geometric parameters of the system as free as possible from systematic
errors, we have made an effort to account for these modulations by
considering up to two spots.

While the subset of the photometric observations used here spans only
22 days, our spectroscopic coverage extends for more than 25 years.
Given the changes in the light curves mentioned in
\S\thinspace\ref{sec:lc_obs}, the photometry is modeled separately
from the radial velocities as any spot model would not apply to
both. In a detached system such as this the photometry does not
constrain the mass ratio, so the value of $q = 0.6773$ was adopted
from the spectroscopy.

The parameters representing the spots in the WD model are the latitude
($\theta_{\rm s}$), longitude ($\phi_{\rm s}$), angular radius
($r_{\rm s}$) and temperature contrast relative to the photosphere
($T_{\rm eff,s}/T_{\rm eff}$). Only cool spots have been considered.
The latitude is essentially unconstrained by our data, and the spot
size and temperature factor are strongly correlated with each other
and cannot usually be adjusted simultaneously. Thus, only the
longitude and the spot size were adjusted at the same time as the
geometric parameters, while the latitude and contrast factor were held
fixed and estimated through a grid search. We explored values over a
wide range in $\theta_{\rm s}$ from +80\arcdeg\ to $-$80\arcdeg\ in
steps of 20\arcdeg, a range in the primary $T_{\rm eff,s}/T_{\rm eff}$
from 0.85 to 0.95, and a range from 0.75 to 0.85 for the secondary
$T_{\rm eff,s}/T_{\rm eff}$, both with a step size of 0.05. As
indicated above, it is virtually impossible to tell which star has the
spot(s), so we have considered here only three possible scenarios: two
spots on the primary, two on the secondary, or one spot on each star.

\begin{deluxetable}{lcc}
\tablewidth{0pc}
\tabletypesize{\scriptsize}
\tablecaption{Light curve parameters of IM~Vir from our simultaneous $BV\!RI$ fit.\label{tab:lc_param}}
\tablehead{\colhead{~~~~~~~~~~~Parameter~~~~~~~~~~~} &
\colhead{Primary} &
\colhead{Secondary}}
\startdata
\multicolumn{3}{l}{Geometric properties\hfil} \\
~~~~Phase shift\dotfill             & \multicolumn{2}{c}{0.0006~$\pm$~0.0001} \\
~~~~$i$ (deg)\dotfill               & \multicolumn{2}{c}{87.24~$\pm$~0.16\phn}     \\
~~~~$\Omega$\dotfill                & 6.298~$\pm$~0.023   & 7.081~$\pm$~0.038    \\
~~~~$r_{pole}$\dotfill              & 0.1776~$\pm$~0.0007 & 0.1145~$\pm$~0.0007  \\
~~~~$r_{point}$\dotfill             & 0.1796~$\pm$~0.0008 & 0.1151~$\pm$~0.0008  \\
~~~~$r_{side}$\dotfill              & 0.1784~$\pm$~0.0007 & 0.1147~$\pm$~0.0007  \\
~~~~$r_{back}$\dotfill              & 0.1776~$\pm$~0.0008 & 0.1150~$\pm$~0.0007  \\
~~~~$r_{\rm vol}$\tablenotemark{a}\dotfill    & 0.1785~$\pm$~0.0008 & 0.1146~$\pm$~0.0010  \\
\multicolumn{3}{l}{Radiative properties\hfil} \\
~~~~$T_{\rm eff}$ (K)\dotfill       & 5570\tablenotemark{b} & 4246~$\pm$~16\phn\phn    \\
~~~~$L_{2}/L_{1}$ $B$ band\dotfill  & \multicolumn{2}{c}{0.04805~$\pm$~0.00008}  \\
~~~~$L_{2}/L_{1}$ $V$ band\dotfill  & \multicolumn{2}{c}{0.07499~$\pm$~0.00016}  \\
~~~~$L_{2}/L_{1}$ $R$ band\dotfill  & \multicolumn{2}{c}{0.10960~$\pm$~0.00027}  \\
~~~~$L_{2}/L_{1}$ $I$ band\dotfill  & \multicolumn{2}{c}{0.14267~$\pm$~0.00036}  \\
~~~~Albedo\tablenotemark{c}\dotfill & \multicolumn{2}{c}{0.5}   \\
~~~~Gravity darkening\tablenotemark{c}\dotfill & 0.34     & 0.40      \\
\multicolumn{3}{l}{Limb darkening coefficients (square root law)\hfil} \\
~~~~$x$ $B$ band\dotfill            & 0.625 & \phs0.904  \\
~~~~$y$ $B$ band\dotfill            & 0.240 & $-$0.057   \\
~~~~$x$ $V$ band\dotfill            & 0.364 & \phs0.644  \\
~~~~$y$ $V$ band\dotfill            & 0.450 & \phs0.190  \\
~~~~$x$ $R$ band\dotfill            & 0.237 & \phs0.462  \\
~~~~$y$ $R$ band\dotfill            & 0.517 & \phs0.354  \\
~~~~$x$ $I$ band\dotfill            & 0.137 & \phs0.269  \\
~~~~$y$ $I$ band\dotfill            & 0.539 & \phs0.475  \\
\multicolumn{3}{l}{Spots properties\hfil} \\
~~~~Latitude\tablenotemark{c} (deg)\dotfill           & $-$60 & $-$60 \\
~~~~Longitude (deg)\dotfill         & 307.3~$\pm$~4.9\phn\phn & 331.8~$\pm$~4.7\phn\phn  \\
~~~~Radius (deg)\dotfill            & 26.4~$\pm$~1.2\phn      & 36.0~$\pm$~2.5\phn \\
~~~~$T_{\rm eff}$ factor\tablenotemark{c}\dotfill    & 0.95 & 0.80 \\
\multicolumn{3}{l}{Residuals and number of observations\hfil} \\
~~~~$\sigma_{B}$ / $N_{B}$\dotfill  & \multicolumn{2}{c}{0.01347 / 438} \\
~~~~$\sigma_{V}$ / $N_{V}$\dotfill  & \multicolumn{2}{c}{0.01295 / 495} \\
~~~~$\sigma_{R}$ / $N_{R}$\dotfill  & \multicolumn{2}{c}{0.01300 / 443} \\
~~~~$\sigma_{I}$ / $N_{I}$\dotfill  & \multicolumn{2}{c}{0.01399 / 455}
\enddata
\tablenotetext{a}{Volume radius derived from the fitted parameters.}
\tablenotetext{b}{Fixed according to the spectroscopic analysis.}
\tablenotetext{c}{Fixed; see text.}
\end{deluxetable}

\begin{figure}[t]
\centering
\includegraphics[width=0.9\columnwidth]{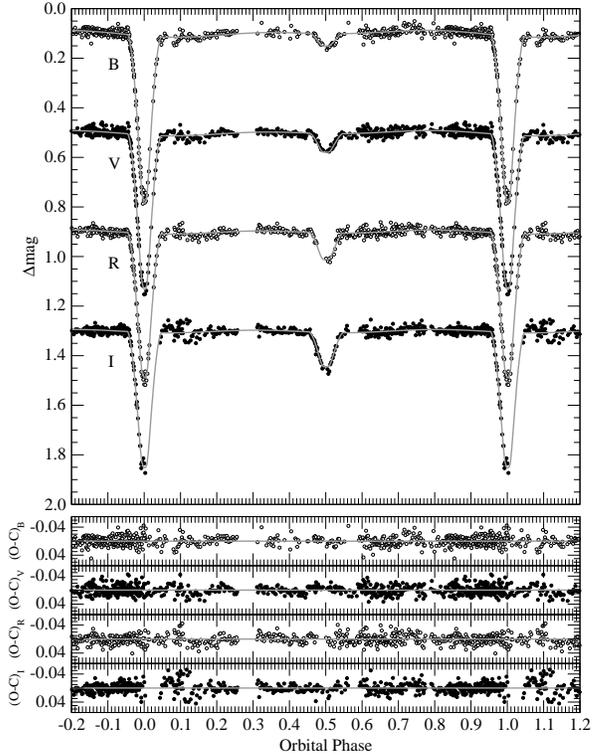}
\caption{$BV\!RI$ observations of IM~Vir along with our best-fit
model.  The curves are shifted vertically for clarity.  Photometric
residuals are shown in the bottom panels, in the same order as the top
curves.}
\label{fig:lc_plot}
\end{figure}

\begin{figure}[t]
\centering
\includegraphics[width=0.9\columnwidth]{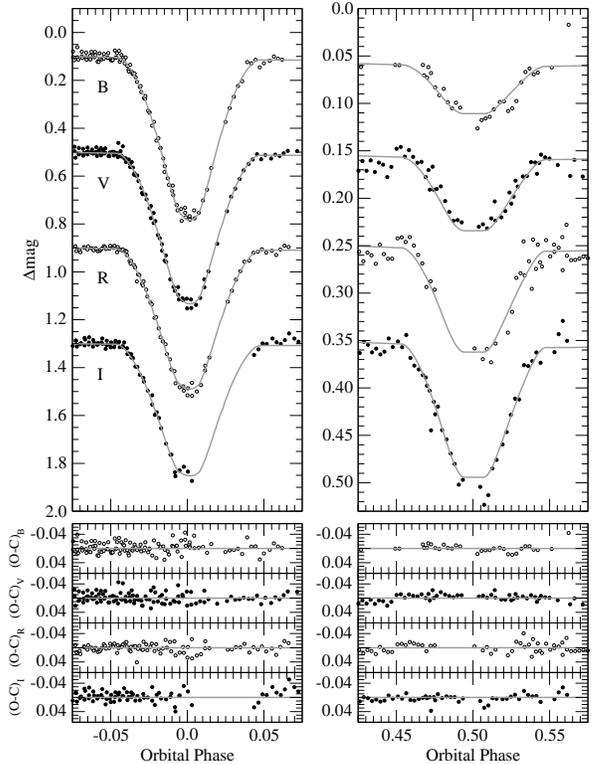}
\caption{Enlargement of the eclipse phases in
Figure~\ref{fig:lc_plot}.  Note the different vertical scales for the
primary and secondary eclipses.}
\label{fig:eclipses_plot}
\end{figure}

\begin{deluxetable*}{l c cccc c ccc c cc}
\tablewidth{0pc}
\tabletypesize{\scriptsize}
\tablecaption{Summary of single-passband light curve fits, and estimated uncertainties.\label{tab:lc_fit}}
\tablehead{\colhead{} & &
\multicolumn{4}{c}{Individual light-curve fits} & &
\multicolumn{3}{c}{Standard error estimates} & &
\multicolumn{2}{c}{Adopted fit} \\
\noalign{\vskip 2pt}
\cline{3-6} 
\cline{8-10} 
\cline{12-13}
\noalign{\vskip 2pt}
\colhead{~~~~Parameter~~~~} & &
\colhead{$B$} & 
\colhead{$V$} &
\colhead{$R$} &
\colhead{$I$} & &
\colhead{$\sigma_{\rm Bands}$} &
\colhead{$\sigma_{\rm Spots}$} &
\colhead{$\sigma_{\rm WD}$} & &
\colhead{Value} &
\colhead{$\sigma$}}
\startdata 
Phase shift\dotfill        &  &   0.0007  &  0.0007  &   0.0006  &   0.0004  & & 0.00014 & 0.00010& 0.00010  && 0.0006 & 0.0002 \\
$i$ (deg)\dotfill          &  &  87.38    & 87.22    &  86.91    &  87.83    & & 0.38    & 0.08   & 0.16     && 87.24  & 0.42   \\
$T_{\rm eff,2}$\dotfill    &  &   4272    &  4191    &   4231    &   4304    & & 49      & 56     & 16       && 4246   &  83    \\
$\Omega_{1}$\dotfill       &  &   6.257   &  6.289   &   6.304   &   6.431   & & 0.077   & 0.018  & 0.023    && 6.298  & 0.082  \\
$\Omega_{2}$\dotfill       &  &   7.091   &  7.080   &   7.003   &   7.220   & & 0.090   & 0.044  & 0.038    && 7.08   & 0.11   \\
$r_{\rm vol,1}$\dotfill    &  &   0.1798  &  0.1788  &   0.1783  &   0.1743  & & 0.0024  & 0.0006 & 0.0008   && 0.1785 & 0.0026 \\
$r_{\rm vol,2}$\dotfill    &  &   0.1144  &  0.1146  &   0.1161  &   0.1119  & & 0.0017  & 0.0009 & 0.0010   && 0.1146 & 0.0022 \\
$L_{2}/L_{1}$ ($B$)\dotfill&  &   0.04999 &  --      &   --      &   --      & & --      & 0.0041 & 0.00021  && 0.0480& 0.0041 \\ 
$L_{2}/L_{1}$ ($V$)\dotfill&  &   --      &  0.05366 &   --      &   --      & & --      & 0.0054 & 0.00030  && 0.0750& 0.0054 \\
$L_{2}/L_{1}$ ($R$)\dotfill&  &   --      &  --      &   0.10928 &   --      & & --      & 0.0065 & 0.00042  && 0.1096& 0.0065 \\
$L_{2}/L_{1}$ ($I$)\dotfill&  &   --      &  --      &   --      &   0.15179 & & --      & 0.0068 & 0.00050  && 0.1427& 0.0068
\enddata
\end{deluxetable*}

\begin{figure}[t]
\centering
\includegraphics[width=\columnwidth]{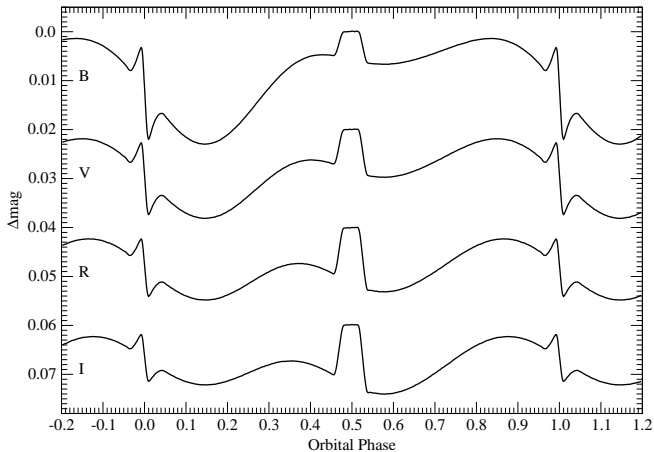}
\caption{Effect of the spots on the light curves. The variations shown
correspond to the adopted scenario with one cool spot on each star.}
\label{fig:spots_LC}
\end{figure}

\begin{figure}[t]
\centering
\includegraphics[width=0.5\columnwidth]{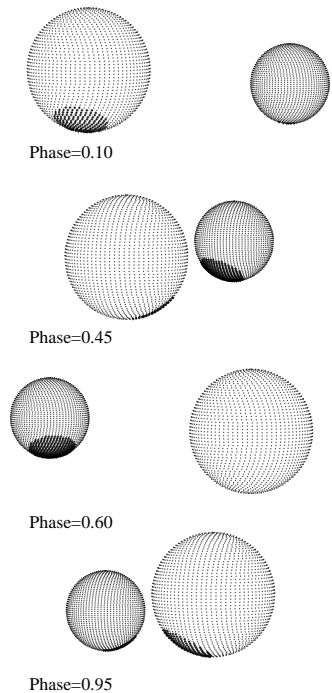}
\caption{Spot location on each star as viewed from the Earth at
different orbital phases, in the adopted scenario in which each
component has one cool spot. The stars and their separation are
rendered to scale.}
\label{fig:spots_map}
\end{figure}

The best solutions were typically obtained with the spots located in
the southern hemisphere, although we do not assign any particular
physical significance to this as we only consider the spot modeling as
a means of removing a perturbation from the underlying light
curve. The overall best fit has one spot on each component, and a
reduced $\chi^2$ square that is only marginally better than the other
two scenarios (1.5\% lower than the case with two spots on the
primary, and 3.4\% lower than the two-spot configuration on the
secondary). The parameters of this best fit are listed in
Table~\ref{tab:lc_param}, and the synthetic curves are shown together
with the observations in Figure~\ref{fig:lc_plot} and
Figure~\ref{fig:eclipses_plot}. Among the quantities derived from this
fit, the light ratios allow for an important consistency check against
the ratio obtained directly from our spectra. Interpolating between
$B$ and $V$ to the mean wavelength of our spectroscopic observations,
we obtain $\ell_2/\ell_1 = 0.066 \pm 0.005$, which agrees well with
the value of $\ell_2/\ell_1 = 0.06 \pm 0.01$ from
\S\thinspace\ref{sec:spectroscopy}.  The effect of the spots on the
light curves is illustrated in Figure~\ref{fig:spots_LC}, and is seen
to range from $\sim$0.025~mag in $B$ to $\sim$0.015~mag in $I$
(peak-to-peak). A depiction of the location of the spots on each star
is shown in Figure~\ref{fig:spots_map}. The spot on the primary covers
5.2\% of its surface, while that of the secondary covers 9.5\%.

In order to provide more realistic uncertainties for the geometric and
radiative parameters than the internal errors reported in
Table~\ref{tab:lc_param}, we have considered additional sources of
error as follows:
(a) We ran WD solutions separately for each
passband, with the spot parameters fixed at their final values, and
examined the differences in the parameters. These fits are summarized
in Table~\ref{tab:lc_fit}. The dispersion in each parameter
about the average of the four bands ($\sigma_{\rm Bands}$)
was taken as an additional contribution to the overall uncertainty.
(b) We calculated the range in each parameter from the simultaneous
$BV\!RI$ fits in the three spot scenarios, and adopted half of this
range as an additional contribution to account for the degeneracy in
the spot modeling.
(c) As a check on the internal errors from WD, we
continued the iterations in our adopted fit beyond convergence for
another 200 steps, and we examined the scatter of those 200
solutions. For the main parameters ($i$, $T_{\rm eff,2}$, $\Omega_1$,
$\Omega_2$) the scatter was only a small fraction of the internal
errors, but for the light ratios it was typically a factor of two
larger. We adopted the larger of the two estimates in each case.

These three sources of error ($\sigma_{\rm Bands}$, $\sigma_{\rm
Spots}$, and $\sigma_{\rm WD}$ in Table~\ref{tab:lc_fit}) were
combined quadratically, and are the ones we assign to the final
light-curve parameters. Changes in the adopted mass ratio within its
error have no additional effect. The parameters and these errors,
which we believe to be realistic, are listed in the last two
columns of Table~\ref{tab:lc_fit}.  Further solutions were carried out
allowing the eccentricity to vary, but in all cases we found the
result to be insignificant compared to its error, consistent with the
indications from spectroscopy.  Third light was also tested for at
various stages of the analysis, but was always found to either
converge toward negative (unphysical) values, or to be consistent with
zero.  Different limb-darkening laws were tested as well (linear,
logarithmic), but the differences with the results in
Table~\ref{tab:lc_fit} were minimal (well within the errors).

The final solution indicates the two stars are nearly spherical, the
difference between $r_{\rm point}$ and $r_{\rm pole}$ being only 1.1\%
for the primary and 0.5\% for the secondary. The primary eclipse is
annular (48\% of the light of that star blocked), while the secondary
is total (see Figure~\ref{fig:eclipses_plot}), with the totality phase
lasting 28 minutes.

\begin{figure}[t]
\centering
\includegraphics[width=\columnwidth]{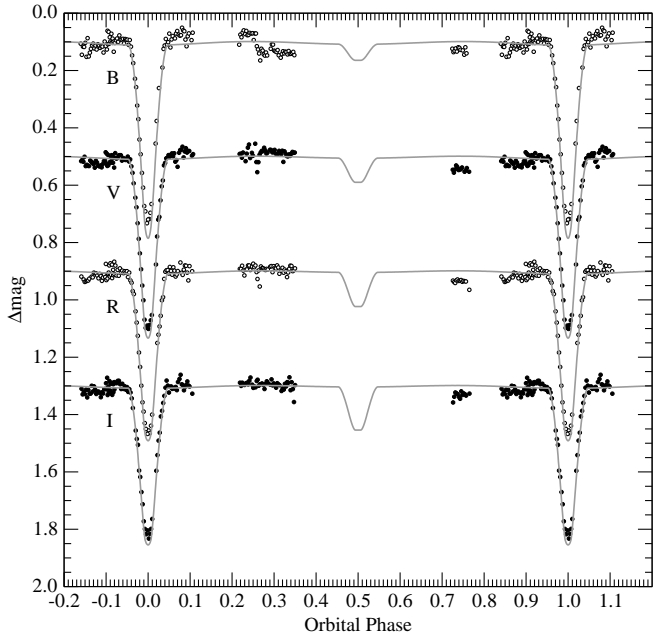}
\caption{Differential $BV\!RI$ photometry of IM~Vir corresponding to
the second half of our data set, which we have not included in the
analysis for the reasons described in the text. The solid curves are
the same best-fit model shown in Figure~\ref{fig:lc_plot} (without
the spots), adjusted for a slight overall brightness change and a phase
shift.}
\label{fig:LC_Season2}
\end{figure}

\begin{figure}[t]
\centering
\includegraphics[width=\columnwidth]{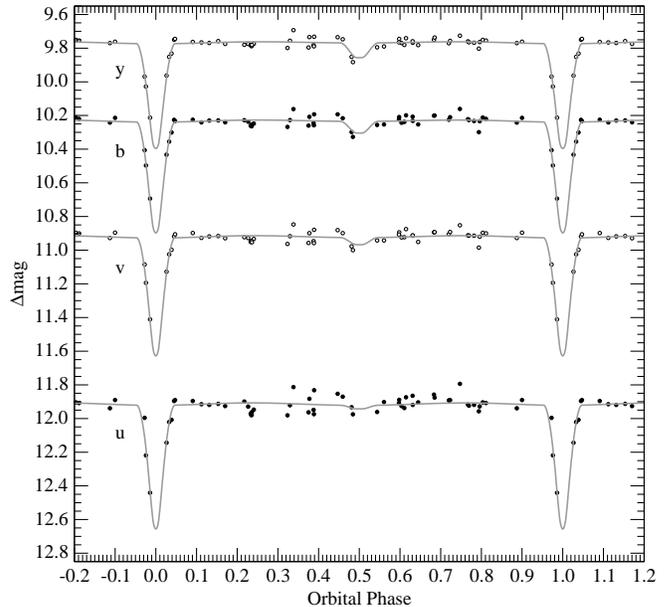}
\caption{Str\"omgren $uvby$ photometry on the standard system
published by \cite{Manfroid1991}, but not used in our analysis,
compared with our best-fit model.}
\label{fig:LC_Manfroid}
\end{figure}

The second half of our photometry, not used here because of likely
changes in the spots and more incomplete phase coverage, is compared
with our final model for the geometry in Figure~\ref{fig:LC_Season2},
excluding the spot terms. The larger scatter is obvious, but the fit
is still quite reasonable. Figure~\ref{fig:LC_Manfroid} displays our
fits to the sparser Str\"omgren photometry by \cite{Manfroid1991}.
Brightness ratios from these solutions (0.0167, 0.0353, 0.0599, and
0.0775 in $uvby$, respectively) are used below to separate the light
of the two components and derive a photometric estimate of their
effective temperatures and metallicity.

\section{Absolute dimensions}
\label{sec:absolute}

\begin{figure}[t]
\centering
\includegraphics[width=0.9\columnwidth]{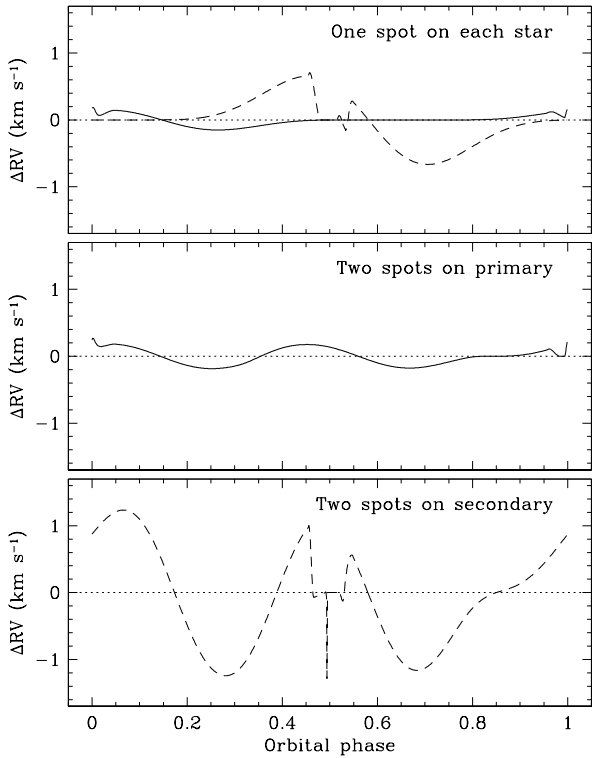}
\caption{Effect of the spots on the radial velocities, for the three
scenarios explored with our WD solutions (solid lines for the primary,
dashed for the secondary).}
\label{fig:spots_RV}
\end{figure}

The absolute masses and radii for the components of IM~Vir follow from
our spectroscopic and light curve solutions described in the preceding
sections. The effect that spots may have on the relative radii has
been considered explicitly in our light-curve modeling, and included
in our error estimates. Similar effects may influence the masses,
although we do not expect them to be significant since they would tend
to average out over the extended coverage of our spectroscopy.
Nevertheless, the spot models in \S\thinspace\ref{sec:lc_analysis}
allow us to quantify the importance of these effects.
Figure~\ref{fig:spots_RV} shows the distortions in the radial
velocities expected from each of the three spot scenarios considered
earlier. As a test, we perturbed the radial velocities using 
these curves and repeated the spectroscopic orbital
solutions. Compared to our adopted fit (Table~\ref{tab:specorbit}),
the differences were less than 0.6\% in the minimum masses. The
differences in $a \sin i$, which affects the absolute radii, were less
than 0.2\%. To be conservative, for the calculation of the final mass
and radius uncertainties we have augmented the errors in
Table~\ref{tab:specorbit} by adding in quadrature half of the range in
$M_{1,2} \sin^3 i$ and $a \sin i$ obtained from the three spot
configurations. With this, the absolute masses of IM~Vir are
determined to 1.2\% and 0.7\% for the primary and secondary, and the
absolute radii to 1.5\% and 1.9\%.

Next in importance to the masses and radii are the effective
temperatures of the stars. The temperature \emph{difference} (or
ratio) is accurately determined from the light curves, but the
absolute scale is set by the primary value, which is fixed in our
analysis. As a check on the spectroscopic value of 5570~K adopted
here, we made use of absolute photometry available for IM~Vir from a
variety of sources, along with color/temperature
calibrations. Brightness measurements were compiled from the 2MASS
catalog \citep{Cutri:03}, the {\it Tycho-2\/} catalog \citep{Hog:00},
Str\"omgren photometry from \cite{Manfroid1991} and \cite{Morale:96},
and our own $BV$ measurements outside of eclipse, transformed to the
standard system using the comparison and check stars. For the
\cite{Manfroid1991} photometry we considered only the data out of
eclipse. The {\it Tycho-2\/} and \cite{Morale:96} measurements are quite
numerous and can be considered to be little affected by the minima.
The date of the 2MASS measurements indicates they were taken during
the egress of a secondary eclipse; corrections of order 0.03~mag
(slightly different in each band) were applied to $JH\!K_s$ based on
our light curve solutions, extended to the near-infrared. The adopted
magnitudes for the combined light in all passbands are listed in
Table~\ref{tab:colors}. With these we formed 8 different color
indices, also listed in the table, and then deconvolved the light to
obtain the individual indices for the two stars. The light ratios in
$B$ and $V$ for the deconvolution were adopted from
Table~\ref{tab:lc_fit}, and those for the Str\"omgren passbands from
our fits in the previous section.  For the other passbands we used
model isochrones (\citealt{Girardi2000} for the {\it Tycho-2\/} bands,
and \citealt{Baraffe1998} for the $JH\!K_s$ bands, after conversion to
the 2MASS system). Two different isochrone ages were tried (1~Gyr and
5~Gyr), although the differences were well within the errors.
Color/temperature calibrations from \cite{Ramirez2005},
\cite{Casagrande2006}, and \cite{GonzalezHernandez2009} were then
applied to the primary indices, assuming solar metallicity. Although
the color indices are not independent of each other, they do serve to
test the internal consistency between the various photometric systems
and the three different calibrations.  The interagreement is very good
(typically within 150~K). The weighted average is $5560 \pm 100$~K
when using the 1~Gyr isochrones for the light ratios, and 40~K less
when adopting 5~Gyr. We regard these to be in very good agreement with
the spectroscopic value, considering the uncertainties. Changing the
metallicity for the calibrations to the rough estimate from
\S\thinspace\ref{sec:spectroscopy} ([m/H] $\approx -0.1$) reduces the
photometric temperatures by less than 10~K. A similar exercise using
the indices for the secondary and an additional calibration for cool
stars by \cite{Casagrande:08} gives a more uncertain result of $4380
\pm 220$~K (as expected from the faintness of the star), but still
consistent with the much more accurate value based on the light curve
analysis, which is $4250 \pm 130$~K. These tests suggest that the
adopted temperatures for IM~Vir are accurate. They correspond to
spectral types of \ion{G7}{5} for the primary and \ion{K7}{5} for the
secondary.

\begin{deluxetable}{lc}
\tablewidth{0pc}
\tablecaption{Out-of-eclipse combined-light magnitudes and colors of IM~Vir.\label{tab:colors}}
\tablehead{
\colhead{~~~~~~Passband~~~~~~} &
\colhead{Value}
}
\startdata
\multicolumn{2}{l}{Magnitudes\hfil} \\
~~~~$B$\dotfill        & 10.234~$\pm$~0.030\phn \\
~~~~$V$\dotfill        &  9.574~$\pm$~0.030 \\
~~~~$b$\dotfill        & 10.209~$\pm$~0.021\phn \\
~~~~$y$\dotfill        &  9.748~$\pm$~0.017 \\
~~~~$B_T$\dotfill      & 10.483~$\pm$~0.039\phn \\
~~~~$V_T$\dotfill      &  9.768~$\pm$~0.030 \\
~~~~$J$\dotfill        &  8.176~$\pm$~0.020\tablenotemark{a} \\
~~~~$H$\dotfill        &  7.712~$\pm$~0.025\tablenotemark{a} \\
~~~~$K_s$\dotfill      &  7.634~$\pm$~0.024\tablenotemark{a} \\
\multicolumn{2}{l}{Color indices\hfil} \\
~~~~$B-V$\dotfill      & 0.661~$\pm$~0.028 \\
~~~~$b-y$\dotfill      & 0.465~$\pm$~0.032 \\
~~~~$V-J$\dotfill      & 1.372~$\pm$~0.028 \\
~~~~$V-H$\dotfill      & 1.832~$\pm$~0.030 \\
~~~~$V-K_s$\dotfill    & 1.905~$\pm$~0.031 \\
~~~~$B_T-V_T$\dotfill  & 0.715~$\pm$~0.057 \\
~~~~$V_T-K_s$\dotfill  & 2.099~$\pm$~0.038 \\
~~~~$J-K_s$\dotfill    & 0.533~$\pm$~0.031
\enddata
\tablenotetext{a}{Corrections of 0.026, 0.030, and 0.035~mag have been
applied to the measured $JH\!K_s$ values to account for the fact that the 2MASS
observation was made during the egress of a secondary eclipse (see
text).}
\end{deluxetable}

An additional quantity of great importance for the interpretation of
the masses, radii, and temperatures is the chemical composition.
Beyond our estimate in \S\thinspace\ref{sec:spectroscopy} of [m/H]
$\approx -0.10 \pm 0.25$, a spectroscopic estimate from a composite
spectrum of IM~Vir was reported by \cite{Dall:07} as [Fe/H] = $-0.53
\pm 0.16$, based on an effective temperature some 200~K cooler and a
higher surface gravity than we derive for the primary, which dominates
the light. It is unclear how accurate this determination is, in view of
those differences.  A photometric estimate for the primary may be derived
from the out-of-eclipse Str\"omgren measurements of IM~Vir by
\cite{Manfroid1991} and \cite{Morale:96}, after removing the light
contribution of the secondary using the light ratios obtained in
\S\thinspace\ref{sec:lc_analysis}.  The metallicity relation by
\cite{Holmberg:07} gives a rather poorly determined value of [Fe/H]
$\approx -0.37 \pm 0.47$, in which the uncertainty includes
photometric errors as well as the scatter of the calibration. The
secondary is too cool for this calibration and other similar ones
based on Str\"omgren indices, but is within range of the near-infrared
formula by \cite{Bonfils:05}, which yields [Fe/H] $\approx -0.26 \pm
0.26$. This again includes all photometric errors and the scatter of
the calibration. It has been noted by \cite{Johnson:09} that this
latter color/temperature relation appears to underestimate the
metallicity of late-type stars of solar composition or greater, by
approximately 0.3~dex; for sub-solar compositions it is not clear that
a correction is necessary. The above estimates suggest a composition
of IM~Vir somewhat below solar, perhaps [Fe/H] $\sim -0.3$, but the
uncertainties are large and this conclusion requires confirmation.

\begin{deluxetable}{lcc}
\tablewidth{0pc}
\tablecaption{Absolute properties of IM~Vir.\label{tab:absolute}}
\tablehead{
\colhead{~~~~~~~~~~Parameter~~~~~~~~~~} &
\colhead{Primary} &
\colhead{Secondary}
}
\startdata
$M$ ($M_{\sun}$)\dotfill         & 0.981~$\pm$~0.012    &   0.6644~$\pm$~0.0048  \\
$R$ ($R_{\sun}$)\dotfill         & 1.061~$\pm$~0.016    &    0.681~$\pm$~0.013  \\
$T_{\rm eff}$ (K)\dotfill        &  5570~$\pm$~100\phn      &     4250~$\pm$~130\phn  \\
$\log g$ (cgs)\dotfill           & 4.379~$\pm$~0.014    &    4.594~$\pm$~0.017  \\
$\log L/L_{\sun}$\dotfill        & $-$0.012~$\pm$~0.034\phs & $-$0.867~$\pm$~0.056\phs  \\
$BC_V$ (mag)\tablenotemark{a}\dotfill    & $-$0.12~$\pm$~0.10\phs   &  $-$0.82~$\pm$~0.17\phs \\
$M_V$ (mag)\tablenotemark{b}\dotfill              &   4.88~$\pm$~0.15   & 7.71~$\pm$~0.29  \\
$v \sin i$ (\kms)\dotfill        &  43~$\pm$~2\phn    &  \nodata  \\
$v_{\rm sync} \sin i$ (\kms)\tablenotemark{c}\dotfill  & 41.0~$\pm$~0.6\phn   &  26.3~$\pm$~0.5\phn  \\
$a$ ($R_{\sun}$)\dotfill         & \multicolumn{2}{c}{5.944~$\pm$~0.020} \\
Distance (pc)\dotfill            & \multicolumn{2}{c}{89.8~$\pm$~5.8\phn}
\enddata
\tablenotetext{a}{The error in these bolometric corrections from
\cite{Flower1996} account for the temperature uncertainties and
include an additional 0.10~mag added in quadrature.}

\tablenotetext{b}{The bolometric magnitude adopted for the Sun is
$M_{\rm bol}^{\sun} = 4.732$, for consistency with the bolometric
corrections.}

\tablenotetext{c}{Projected rotational velocity if synchronized with
the orbital motion.}

\end{deluxetable}

The absolute dimensions for the system are summarized in
Table~\ref{tab:absolute}, along with derived properties including the
luminosities and absolute visual magnitudes. To calculate $M_V$ we
have adopted the bolometric corrections from \cite{Flower1996}, with
uncertainties that account for the temperature errors as well as a
contribution of 0.10~mag added in quadrature, to be
conservative. IM~Vir does not have an entry in the {\it Hipparcos\/}
Catalog \citep{Perryman:97}. Here we estimate the distance to the
system to be $89.8 \pm 5.8$~pc, ignoring extinction for such a close
object. This is 50\% larger than first estimated by
\cite{Strassmeier1993} on the basis of the spectral type. Separate
distance estimates for each component agree nearly perfectly, showing
the internal consistency of the fundamental data on which they are
based. The distance, proper motions from {\it Tycho-2\/}, and systemic
velocity from our spectroscopic orbital solution
(Table~\ref{tab:specorbit}) lead to space velocities of $U =
+24.4$~\kms, $V = -17.0$~\kms, and $W = -3.0$~\kms\ in the solar frame
(with $U$ positive toward the Galactic center). These do not appear to
associate IM~Vir with any known moving group in the solar
neighborhood.

With our accurate radii we compute projected synchronous rotational
velocities of $41.0 \pm 0.6$~\kms\ and $26.3 \pm 0.5$~\kms\ for the
primary and secondary, respectively. The value for the primary is
consistent with our direct measurement of $43 \pm 2$~\kms, suggesting
that synchronization with the orbital motion has been achieved (if the
spin axis is parallel to the orbital axis). This is expected from the
short period.  Estimates of 43~\kms\ for the primary and 23~\kms\ for
the secondary were reported by \cite{Strassmeier1993}, but without
errors.  At our request, those spectra were kindly remeasured by F.\
Fekel (priv.\ comm.), giving $42 \pm 2$~\kms\ and $31 \pm 4$~\kms,
based on the calibration by \cite{Fekel:97}. Once again the primary
agrees well with the synchronous value, and the secondary is probably
also consistent, considering the difficulty of the measurement.  An
independent value for the primary was reported by
\cite{Strassmeier2000} as 36.2~\kms\ with an uncertainty of 2--4~\kms,
roughly in agreement with ours.

\section{Discussion}
\label{sec:discussion}

With its solar-type primary and low-mass secondary, IM~Vir is a
particularly interesting system for testing models of stellar
evolution. The accurate masses, radii, and temperatures for both
components offer an opportunity to further investigate the
discrepancies for low-mass stars mentioned in the Introduction.  The
leverage afforded by the very different masses is unique among systems
with at least one component under 0.8~$M_{\sun}$.

\begin{figure}[t]
\centering
\includegraphics[width=\columnwidth]{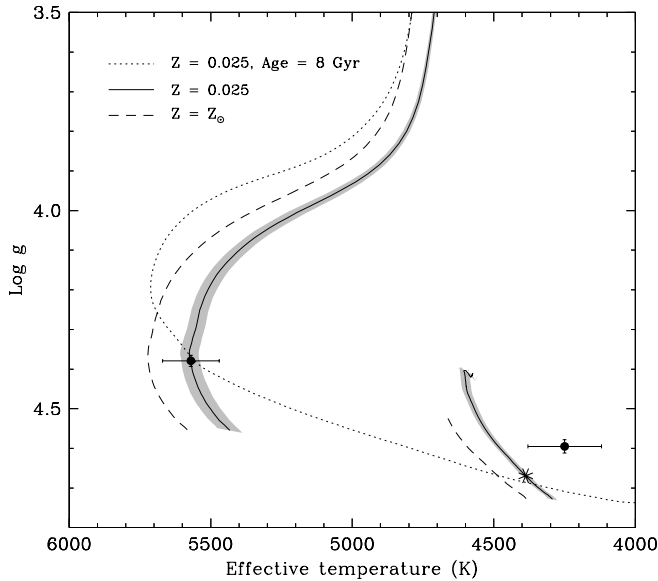}
\caption{Evolutionary tracks for the measured masses of the components
of IM~Vir from the Yonsei-Yale series \citep{Yi:01,
Demarque2004}. Dashed lines correspond to solar metallicity, solid
lines to the composition of $Z = 0.025$ that fits the primary
best. The mass errors are indicated by the shaded areas. The implied
age for the primary is 8~Gyr, and the corresponding isochrone is shown
as the dotted line. The track for the secondary is calculated through
the end of the main-sequence phase, which is reached at an age of
$\sim$40~Gyr. The age of the universe is marked with an asterisk on
this track. The observed location of the secondary would thus point to
an implausibly old age for the star. This is a result of the models
overestimating the temperature and underestimating the size for this
star.}
\label{fig:yaletracks}
\end{figure}

We begin by comparing the measured properties against models from the
widely used Yonsei-Yale series \citep{Yi:01, Demarque2004}, which
treat convection in the standard mixing-length approximation with a
mixing length parameter $\alpha_{\rm ML} = 1.7432$ (in units of the
pressure scale height), calibrated against the Sun. Evolutionary
tracks for solar composition and for the exact masses we measure for
each star are shown in the $\log g$--$T_{\rm eff}$ diagram of
Figure~\ref{fig:yaletracks}, as dashed lines. These models are seen to
be too hot compared to the estimated temperatures.  Adjusting the
metallicity to higher values brings the tracks closer to the observations,
but it is not possible to match both stars at the same time, especially
considering that the temperature \emph{difference} is much better
known than the absolute temperatures. Part of
the problem has to do with the fact that these models are not intended
for low-mass stars such as IM~Vir~B, which require a more
sophisticated equation of state, and particularly, non-gray boundary
conditions between the interior and the photosphere \citep[see,
e.g.,][]{Chabrier:97}. If we therefore focus for the moment only on
the solar-type primary, we find an excellent match to an evolutionary
track for $Z = 0.025$ (corresponding to [Fe/H] $= +0.15$ in these
models), at the rather old age of about 8~Gyr. The corresponding mass
tracks for both stars are drawn with solid lines in
Figure~\ref{fig:yaletracks}, and the shaded area indicates the
uncertainty in their location that comes from our $\sim$1\% mass
errors. An isochrone for this age is indicated with the dotted line.

\begin{figure}[t]
\centering
\includegraphics[width=\columnwidth]{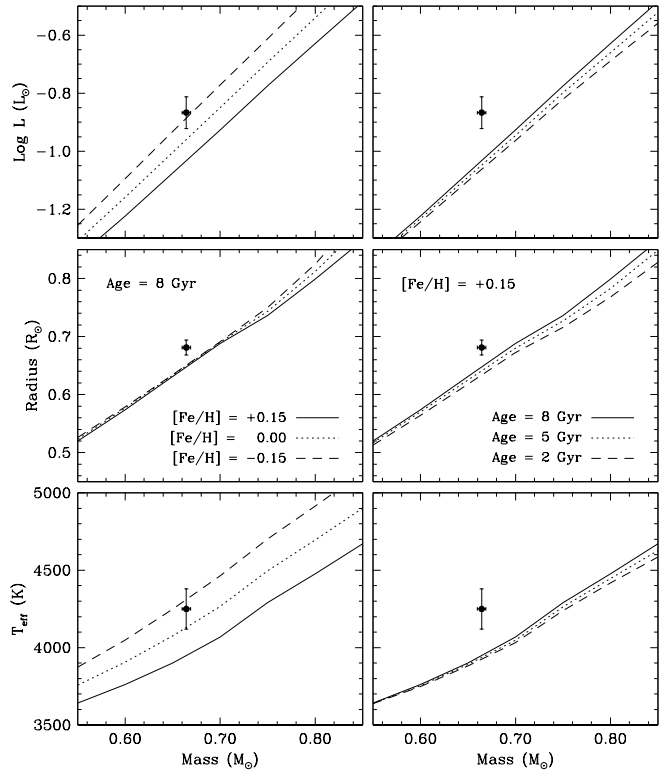}
\caption{Secondary properties compared against \cite{Baraffe1998}
models for parameters that best fit the primary star, when using the
Yonsei-Yale models (see text and Figure~\ref{fig:yaletracks}).
{\it Left panels:} Fixed age of 8 Gyr, and different metallicities,
as labeled. {\it Right panels:} Fixed abundance of [Fe/H] = +0.15, and
different ages, as labeled.}
\label{fig:testyale}
\end{figure}

Not surprisingly, the secondary does not fit the evolutionary track at
this metallicity, and appears implausibly old. This is a consequence
of it being both too large and too cool compared to the models, which
is in the same sense as deviations found for many other late-type
dwarfs, and is believed to be due to the effects of chromospheric
activity.  Previous studies have indicated that better agreement for
this class of stars is possible with models such as those of
\cite{Baraffe1998}, which have the required non-gray boundary
conditions and use a lower mixing-length parameter of $\alpha_{\rm ML}
= 1.0$. The latter tends to diminish (but does not completely
eliminate) the discrepancies noted with other models in $R$ and
$T_{\rm eff}$. The one property of low-mass stars that appears to be
reasonably well reproduced by theory is the luminosity \citep[see,
e.g.,][]{Delfosse:00, Torres:02}. This provides a means of testing the
predictions from the Yonsei-Yale models for the primary, which point
to an unexpected combination of super-solar metallicity and old age,
and a composition that is somewhat inconsistent with the indications
from \S\thinspace\ref{sec:absolute}.  In the left panels of
Figure~\ref{fig:testyale} we compare the secondary luminosity, radius,
and temperature against isochrones from \cite{Baraffe1998}
corresponding to the age of 8~Gyr found above, for different
compositions including the value [Fe/H] $= +0.15$ that best fits the
primary (according to the Yonsei-Yale models).\footnote{For this last
isochrone we have extrapolated slightly in metallicity from the
publicly available tables by \cite{Baraffe1998} for [Fe/H] = 0.0 and
$-0.5$.} All three of $\log L$, $R$, and $T_{\rm eff}$ for the
secondary are seen to be too large compared to the model favored by
the primary, which is represented by the solid lines. The measured
luminosity of the secondary, as well as its effective temperature,
suggest a significantly lower abundance, and the measured radius is
too large no matter what the metallicity. Changes in age at a fixed
metallicity of [Fe/H] = +0.15 do not improve the fit, as seen clearly
in the right panels of Figure~\ref{fig:testyale}.

There is little doubt that the radius of the secondary is too large
compared to models.  Furthermore, the age/metallicity inconsistency
between the primary and secondary is perhaps an indication that the
good match for IM~Vir~A in Figure~\ref{fig:yaletracks} is illusory,
and that it too may have anomalies similar to those of the secondary,
possibly related also to activity. The primary is, after all, a rapid
rotator ($v \sin i = 43$~\kms). Standard models such as those of
\cite{Baraffe1998} do not account for the effects of chromospheric
activity, although the artificially low value of the mixing length
parameter ($\alpha_{\rm ML} = 1.0$) seems to be a step in the right
direction \citep[see also][]{Torres:06, Chabrier:07, Clausen:09}. 

\begin{figure}[t]
\centering
\includegraphics[width=0.45\columnwidth]{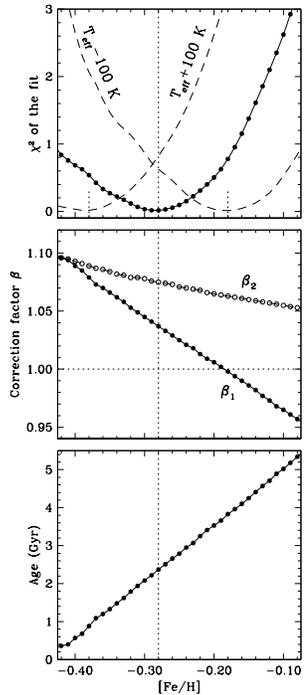}
\caption{Illustration of a grid search to obtain the best simultaneous
match to the measured properties of the primary and secondary of
IM~Vir using the \cite{Baraffe1998} models. These models are adjusted
here by including correction factors $\beta_1$ and $\beta_2$ to the
theoretical radii to account for the effects of chromospheric activity
(see text). {\it Top:} The $\chi^2$ of the fit as a function of metallicity
for the nominal effective temperatures of the stars (solid line with filled
circles), and for temperatures perturbed by $\pm$100~K to explore the role of
systematics (dashed lines). {\it Middle:} Correction factors to the
predicted radii. {\it Bottom:} Best-fit age as a function of
metallicity.}
\label{fig:betagrid}
\end{figure}

\begin{figure}[t]
\centering
\includegraphics[width=0.9\columnwidth]{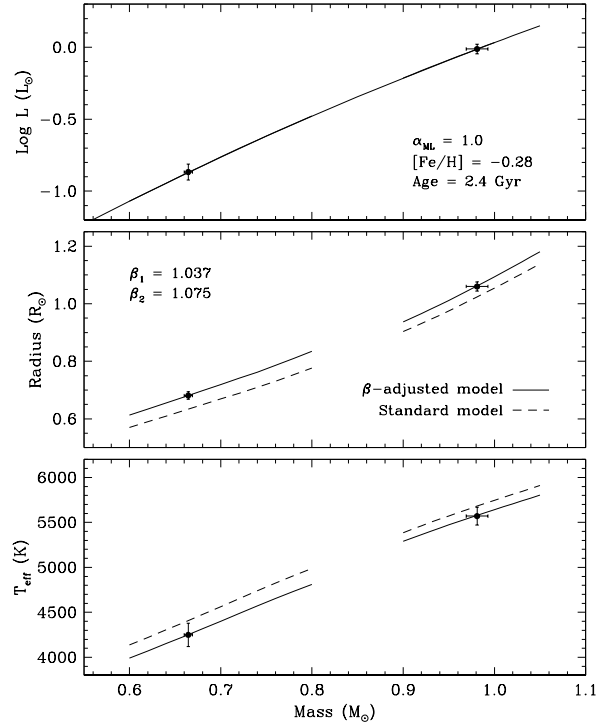}
\caption{Best-fit model from \cite{Baraffe1998} for [Fe/H] $= -0.28$
and an age of 2.4~Gyr (solid lines), with the radii adjusted by the
correction factors $\beta_1$ and $\beta_2$ for the primary and
secondary, respectively (see Figure~\ref{fig:betagrid} and text). The
theoretical temperatures are adjusted by factors $\beta^{-1/2}$, and
the model luminosity is unchanged. The dashed lines in the two bottom
panels represent the standard models for the same age and metallicity,
but without any corrections for activity ($\beta_{1,2} = 1.000$).}
\label{fig:betamodel}
\end{figure}

As a way of parameterizing the missing physical effects from activity
and their impact on the structure of low-mass stars, \cite{Torres:07}
explored the use of a correction factor $\beta$ to the theoretical
radii, and showed that good fits to empirical data could be achieved
by simultaneously correcting the theoretical temperatures by
$\beta^{-1/2}$, so as to preserve the bolometric luminosity. We apply
the same procedure here. To account for a possible difference in the
activity level, we have considered separate values of $\beta$ for each
star, and examined a wide range of metallicities and ages using the
\cite{Baraffe1998} models to obtain the best simultaneous match to the
properties of both components, at a single age and composition. In
Figure~\ref{fig:betagrid} we display the results of our grid search. A
near-perfect fit is achieved for [Fe/H] $= -0.28$, as measured by the
$\chi^2$ represented in the top panel. Interestingly, this value is
much more consistent with the rough estimates from photometry and
spectroscopy described in \S\thinspace\ref{sec:absolute} than with the
metal-rich composition suggested by the Yonsei-Yale models for the primary.
The radius correction factors are $\beta_1 = 1.037$ and $\beta_2 = 1.075$,
and the best-fit age is 2.4~Gyr (see middle and bottom panels). The age
and metallicity depend almost entirely on the measured stellar
luminosities. The values of $L$, $R$, and $T_{\rm eff}$ are shown in
Figure~\ref{fig:betamodel} with the adjusted isochrones from
\cite{Baraffe1998}. Standard models without the corrections to the
radii and temperatures are shown as dashed lines in the lower panels,
for reference.

The implication of the above fit is that the radius of the primary
star is 3.7\% larger than predicted by theory, and its temperature is
just under 2\% (100~K) too cool. For the secondary the models
underpredict the radii by 7.5\%, and overpredict the temperature by
about 3.5\%, or 150~K. It appears, therefore, that \emph{both} stars
are affected by chromospheric activity. While this is not unexpected
for the secondary, the fact that the primary also seems to show the
same anomalies supports other recent evidence that the phenomenon is
not confined to low-mass stars, but reaches stars of solar mass
(0.981~$M_{\sun}$ for IM~Vir~A). Other examples of solar-type stars
where this has been documented include VZ~Cep~B
\citep[1.108~$M_{\sun}$;][]{Torres:09}, CV~Boo~B
\citep[0.968~$M_{\sun}$;][]{Torres:08}, FL~Lyr~B
\citep[0.958~$M_{\sun}$;][]{Popper:86}, and V1061~Cyg~B
\citep[0.932~$M_{\sun}$;][]{Torres:06}.

There is ample evidence of activity in IM~Vir, which was in fact
discovered through its X-ray emission, as described in
\S\thinspace\ref{sec:introduction}. Activity is manifested in our
light curves in the form of spottedness, and in retrospect, our
finding that the best fit to the photometry is achieved with one spot
on each star as opposed to spots on any single component is consistent
with the results of the stellar evolution modeling in the preceding
paragraph, which indicates that both stars are affected. This is not
surprising, given that both have convective envelopes and are rapid
rotators. Additional manifestations of activity in this case include
a filled-in H$\alpha$ line \citep{Strassmeier1993, Liu:96} or the possible
detection of H$\alpha$ emission from the secondary \citep{Popper:96},
\ion{Ca}{2}~H and K emission \citep{Dall:07}, and flaring in X~rays
\citep{Pandey2008}.  With the X-ray count rates and hardness ratio
from \textit{ROSAT}, the energy conversion factor prescribed by
\cite{Schmitt:95}, and our bolometric luminosities and derived
distance, we have estimated the ratio of the X-ray to bolometric
luminosity for the two components. We obtain $\log(L_{\rm X}/L_{\rm
bol}) = -3.70 \pm 0.13$ and $-2.84 \pm 0.13$, respectively, where we
have assumed equal X-ray emission from each star since the
\textit{ROSAT} observation does not resolve the binary. These values
are consistent with the secondary being completely saturated, and the
primary being near saturation. They are similar to the X-ray
luminosities seen in other active binary systems, and are in fact in
good agreement with the trend between the radius anomalies and $L_{\rm
X}/L_{\rm bol}$ found by \cite{LopezMorales2007} (see, e.g., her
Figure~2).

The differences in radius and temperature between the models and the
observations for IM~Vir are of similar magnitude as those found for
other active stars with convective envelopes, but do rely quite
strongly on our adopted effective temperatures for the stars because
of the way the $\beta$ correction factors have been determined. Given
the greater difficulty of determining absolute temperatures
than other properties such as masses and radii, one
may wonder to what extent systematic errors in $T_{\rm eff}$ might
affect the result, and also the derived age and metallicity. We focus
here on systematic errors in the primary temperature, since the
secondary value is tied to the primary through the light-curve
solution.  We explored this question by perturbing the temperatures by
$\pm$100~K, with corresponding adjustments to the luminosities, and
repeating the grid search for $\beta$. The results are illustrated in
the top panel of Figure~\ref{fig:betagrid}, where the dashed lines
show that the best-fit metallicity changes by $\pm$0.10~dex (marked by
the short dotted lines).  The changes in the $\beta$ factors are
$\pm$0.010 for $\beta_1$ and $\pm$0.003 for $\beta_2$, and the change
in the best-fit age is $\pm$0.50~Gyr, or $\sim$20\%.

\section{Concluding remarks}
\label{sec:conclusions}

Our extensive spectroscopic and photometric observations of IM~Vir
have enabled us to determine highly accurate values for the absolute
masses and radii of both stars to better than 2\%, as well as accurate
temperatures (Table~\ref{tab:absolute}). This eclipsing system now
joins the ranks of those with the best determined properties
\citep[see][]{Torresetal:09}. The primary is of solar type, and the
secondary is a late K dwarf. The very different masses provide
increased discriminating power for testing models of stellar
evolution. We find evidence that both components, which have
convective envelopes and are rapidly rotating, show discrepancies in
their radii and temperatures compared to calculations by
\cite{Baraffe1998}, similar to those reported for other low-mass
stars. The predicted radii are too small by 3.7\% and 7.5\% for the
primary and secondary, and the theoretical temperatures are too high
by 100~K and 150~K, respectively. These effects are ascribed to
chromospheric activity, for which there is abundant evidence in this
system in the form of X-ray, \ion{Ca}{2}~H and K, and H$\alpha$
emission, X-ray flaring, and spottedness seen directly in the light
curves. The larger effect observed for the secondary is consistent
with the larger fraction of that star being covered by spots (a factor
of two difference). The fact that the near-solar mass primary is also
affected supports recent findings for other binaries indicating that
the impact of activity on the structure of stars is not limited to the
M dwarfs, as previously thought, but most likely extends to all stars
with convective envelopes, reaching up to masses near or larger than that
of the Sun.  Current stellar evolution models do not account for these
effects. Some progress in this area has been made to incorporate
magnetic activity in the theoretical calculations
\citep[e.g.,][]{Dantona:00, Mullan:01, Chabrier:07}, and initial
results are very encouraging.

A crucial ingredient for the comparison with models is the chemical
composition. The best-fit models for IM~Vir indicate an age of $2.4
\pm 0.5$~Gyr and a metallicity of [Fe/H] $= -0.28 \pm 0.10$. The
uncertainties here represent the sensitivity to systematic errors of
$\pm$100~K in the effective temperatures, but exclude unquantified
systematics in the \cite{Baraffe1998} models themselves, which are
difficult to evaluate. While the metallicity prediction seems to be in
good agreement with our rough spectroscopic abundance estimate in
\S\thinspace\ref{sec:spectroscopy} and that of \cite{Dall:07}, as well
as photometric estimates for both stars, a proper detailed
spectroscopic analysis of both components is still lacking, and is
essential to validate the comparison.  Such determinations are
notoriously difficult in double-lined binaries, which is why
relatively few have them. In IM~Vir this task would be challenging for
two reasons: the secondary is very faint compared to the primary, and
it is of late spectral type. Abundance determinations for late-type
stars are still problematic due to shortcomings in the model
atmospheres. However, IM~Vir offers a unique opportunity because the
secondary eclipse is total. High S/N ratio spectra taken with a
sufficiently large telescope during the 28-minute totality phase would
be of the primary only, and can be analyzed with standard techniques.
The uncertainty in this observing window is estimated to be approximately
$\pm$5 minutes, based on the uncertainties in the geometric light-curve
parameters.

\acknowledgements

We are grateful to I.\ Ribas and C.\ Jordi for useful discussions
on the implementation of the Wilson-Devinney code. We also thank F.\
Fekel for communicating his $v \sin i$ measurements of IM~Vir, and
the anonymous referee for helpful suggestions. The
spectroscopic observations used in this work were obtained with
the able help of P.\ Berlind, M.\ Calkins, J.\ Caruso, G.\ Esquerdo,
R.\ Davis, E.\ Horine, D.\ Latham, R.\ Mathieu, D.\ Silva, J.\
Stauffer, R.\ Stefanik, S.\ Tokarz, and J.\ Zajac.  We also thank R.\
Davis for maintaining the CfA echelle database. JCM
acknowledges financial support from the Spanish Ministerio de Ciencia
e Innovaci\'on during the research stay at the Harvard-Smithsonian
Center for Astrophysics, where most of this work was done, and from
the Spanish Ministerio de Educaci\'on y Ciencia via grants
AYA2006-15623-C02-01 and AYA2006-15623-C02-02. GT acknowledges
partial support for this work from NSF grant AST-0708229.
LM and WB were supported by Gettysburg College. Additional thanks go
to David Kraft for preliminary data reductions, and to Peter Mack and
Garry Hummer for technical support at the Gettysburg College
Observatory.  This research has made use of the SIMBAD database and
the VizieR catalogue access tool, both operated at CDS, Strasbourg,
France, of NASA's Astrophysics Data System Abstract Service, and of
data products from the Two Micron All Sky Survey (2MASS), which is a
joint project of the University of Massachusetts and the Infrared
Processing and Analysis Center/California Institute of Technology,
funded by NASA and the NSF. We also made use of data provided by the
VSOP collaboration, through the VSOP wiki database operated at ESO
Chile and ESO Garching.


\begin{thebibliography}{}
\bibitem[Andersen(1991)]{Andersen1991} Andersen, J.\ 1991, \aapr, 3, 91
\bibitem[Baraffe et al.(1998)]{Baraffe1998} Baraffe, I., Chabrier, G., Allard, F., \& Hauschildt, P.\ H.\ 1998, \aap, 337, 403
\bibitem[Bonfils et al.(2005)]{Bonfils:05} Bonfils, X., Delfosse, X., Udry, S., Santos, N.\ C., Forveille, T., \& S\'egransan, D. 2005, \aap, 442, 635
\bibitem[Casagrande et al.(2006)]{Casagrande2006} Casagrande, L., Portinari, L., \& Flynn, C.\ 2006, \mnras, 373, 13
\bibitem[Casagrande et al.(2008)]{Casagrande:08} Casagrande, L., Flynn, C., \& Bessell, M. 2008, \mnras, 389, 585
\bibitem[Chabrier et al.(2007)]{Chabrier:07} Chabrier, G., Gallardo, J., \& Baraffe, I. 2007, \aap, 472, L17
\bibitem[Chabrier \& Baraffe(1997)]{Chabrier:97} Chabrier, G., \& Baraffe, I. 1997, \aap, 327, 1039
\bibitem[Claret(2000a)]{Claret2000a} Claret, A.\ 2000a, \aap, 363, 1081
\bibitem[Claret(2000b)]{Claret2000b} Claret, A.\ 2000b, \aap, 359, 289
\bibitem[Clausen et al.(2008)]{Clausen:08} Clausen, J.\ V., Torres, G., Bruntt, H., Andersen, J., Nordstr\"om, B., Stefanik R.\ P., Latham, D.\ W., \& Southworth, J. 2008, \aap, 487, 1095
\bibitem[Clausen et al.(2009)]{Clausen:09} Clausen J.\ V., Bruntt H., Claret A., Larsen A., Andersen J., Nordstr\"om B., Gim\'enez A. 2009, \aap, 502, 253
\bibitem[Cutri et al.(2003)]{Cutri:03} Cutri, R.\ M.\ et al.\ 2003, 2MASS All-Sky Catalog of Point Sources, Univ.\ of Massachusetts and Infrared Processing and Analysis Center (IPAC/California Institute of Technology)
\bibitem[Dall et al.(2007)]{Dall:07} Dall, T.\ H.\ et al.\ 2007, \aap, 470, 1201
\bibitem[D'Antona et al.(2000)]{Dantona:00} D'Antona F., Ventura P., \& Mazzitelli I. 2000, \apjl, 543, L77
\bibitem[Delfosse et al.(2000)]{Delfosse:00} Delfosse, X., Forveille, T., S\'egransan, D., Beuzit, J.-L., Udry,S., Perrier, C., \& Mayor, M. 2000, \aap, 364, 217
\bibitem[Demarque et al.(2004)]{Demarque2004} Demarque, P., Woo, J.-H., Kim, Y.-C., \& Yi, S.\ K.\ 2004, \apjs, 155, 667
\bibitem[Eker(1996)]{Eker:96} Eker, Z. 1996, \apj, 473, 388
\bibitem[Eker(1999)]{Eker:99} Eker, Z. 1999, \apj, 512, 386
\bibitem[Fekel(1997)]{Fekel:97} Fekel, F.\ C. 1997, \pasp, 109, 514
\bibitem[Flower(1996)]{Flower1996} Flower, P.\ J.\ 1996, \apj, 469, 355
\bibitem[Girardi et al.(2000)]{Girardi2000} Girardi, L., Bressan, A., Bertelli, G., \& Chiosi, C.\ 2000, \aaps, 141, 371
\bibitem[Gonz\'alez Hern\'andez \& Bonifacio(2009)]{GonzalezHernandez2009} Gonz\'alez Hern\'andez, J.\ I., \& Bonifacio, P.\ 2009, \aap, 497, 497
\bibitem[Griffin et al.(2000)]{Griffin:00} Griffin, R.\ E.\ M., David, M., \& Verschueren, W. 2000, \aaps, 147, 299
\bibitem[Helfand \& Caillault(1982)]{Helfand1982} Helfand, D.\ J., \& Caillault, J.-P.\ 1982, \apj, 253, 760
\bibitem[H\o g et al.(2000)]{Hog:00} H\o g, E., Fabricius, C., Makarov, V.\ V., Urban, S., Corbin, T., Wycoff, G., Bastian, U., Schwekendiek, P., \& Wicenec, A. 2000, \aap, 355, L27
\bibitem[Holmberg et al.(2007)]{Holmberg:07} Holmberg, J., Nordstr\"om, B., \& Andersen, J. 2007, \aap, 475, 519
\bibitem[Johnson \& Apps(2009)]{Johnson:09} Johnson, J.\ A., \& Apps, K. 2009, \apj, 699, 933
\bibitem[Lacy et al.(2008)]{Lacy:08} Lacy, C.\ H.\ S., Torres, G., \& Claret, A. 2008, \aj, 135, 1757
\bibitem[Latham(1985)]{Latham:85} Latham, D.\ W. 1985, in IAU Coll.\ 88, Stellar Radial Velocities, eds.\ A.\ G.\ D.\ Philip \& D.\ W.\ Latham (Schenectady: L.\ Davis), 21
\bibitem[Latham(1992)]{Latham:92} Latham, D.\ W. 1992, in IAU Coll.\ 135, Complementary Approaches to Double and Multiple Star Research, ASP Conf.\ Ser.\ 32, eds.\ H.\ A.\ McAlister \& W.\ I.\ Hartkopf (San Francisco: ASP), 110
\bibitem[Latham et al.(1996)]{Latham:96} Latham, D.\ W., Nordstr\"om, B., Andersen, J., Torres, G., Stefanik, R.\ P., Thaller, M., \& Bester, M. 1996, \aap, 314, 864
\bibitem[Latham et al.(2002)]{Latham:02} Latham, D.\ W., Stefanik, R.\ P., Torres, G., Davis, R.\ J., Mazeh, T., Carney, B.\ W., Laird, J.\ B., \& Morse, J.\ A. 2002, \aj, 124, 1144
\bibitem[Liu et al.(1996)]{Liu:96} Liu, X., Huang, H., \& Zhu, W. 1996, \apss, 246, 39
\bibitem[L\'opez-Morales(2007)]{LopezMorales2007} L\'opez-Morales, M.\ 2007, \apj, 660, 732
\bibitem[L\'opez-Morales \& Ribas(2005)]{LopezMorales2005} L\'opez-Morales, M., \& Ribas, I.\ 2005, \apj, 631, 1120
\bibitem[Manfroid et al.(1991)]{Manfroid1991} Manfroid, J. et al.\ 1991, \aaps, 87, 481
\bibitem[Marschall et al.(1988)]{Marschall1988} Marschall, L.\ A., Nations, H., \& Witman, K.\ 1988, \baas, 20, 994
\bibitem[Marschall et al.(1989)]{Marschall1989} Marschall, L.\ A., Stefanik, R., Nations, H., \& Karshner, G.\ 1989, \baas, 21, 1083
\bibitem[Morale et al.(1996)]{Morale:96} Morale, F., Micela, G., Favata, F., \& Sciortino, S. 1996, \aaps, 119, 403
\bibitem[Morales et al.(2009)]{Morales2009} Morales, J.\ C., et al.\ 2009, \apj, 691, 1400
\bibitem[Mullan \& MacDonald(2001)]{Mullan:01} Mullan D.\ J., \& MacDonald, J. 2001, \apj, 559, 353
\bibitem[Ogloza et al.(2008)]{Ogloza2008} Ogloza, W., Niewiadomski, W., Barnacka, A., Biskup, M., Malek, K., \& Sokolowski, M.\ 2008, IBVS, 5843, 1
\bibitem[Pandey \& Singh(2008)]{Pandey2008} Pandey, J.\ C., \& Singh, K.\ P.\ 2008, \mnras, 387, 1627
\bibitem[Perryman et al.(1997)]{Perryman:97} Perryman, M.\ A.\ C., et al.\ 1997, The {\it Hipparcos\/} and {\it Tycho\/} Catalogues (ESA SP-1200; Noordwjik: ESA)
\bibitem[Popper(1997)]{Popper:96} Popper, D.\ M. 1996, \apjs, 106, 133
\bibitem[Popper(1997)]{Popper:97} Popper, D.\ M. 1997, \aj, 114, 1195
\bibitem[Popper(2000)]{Popper:00} Popper, D.\ M. 2000, \aj, 119, 2391
\bibitem[Popper et al.(1986)]{Popper:86} Popper, D.\ M., Lacy, C.\ H., Frueh, M.\ L., \& Turner, A.\ E. 1986, \aj, 91, 383
\bibitem[Ram\'irez \& Mel\'endez(2005)]{Ramirez2005} Ram\'irez, I., \& Mel\'endez, J.\ 2005, \apj, 626, 446
\bibitem[Ribas(2003)]{Ribas2003} Ribas, I. 2003, \aap, 398, 239
\bibitem[Ribas et al.(2008)]{Ribas2008} Ribas, I., Morales, J.~C., Jordi, C., Baraffe, I., Chabrier, G., \& Gallardo, J.\ 2008, \memsai, 79, 562
\bibitem[Schmitt et al.(1995)]{Schmitt:95} Schmitt, J.\ H.\ M.\ M., Fleming, T.\ A., \& Giampapa, M.\ S. 1995, \apj, 450, 392
\bibitem[Siess et al.(2000)]{Siess2000} Siess, L., Dufour, E., \& Forestini, M.\ 2000, \aap, 358, 593
\bibitem[Silva et al.(1987)]{Silva1987} Silva, D.\ R., Gioia, I.\ M., Maccacaro, T., Mereghetti, S., \& Stocke, J.\ T.\ 1987, \aj, 93, 869
\bibitem[Strassmeier et al.(1993)]{Strassmeier1993} Strassmeier, K.\ G., Hall, D.\ S., Fekel, F.\ C., \& Scheck, M.\ 1993, \aaps, 100, 173
\bibitem[Strassmeier et al.(2000)]{Strassmeier2000} Strassmeier, K.\ G., Washuettl, A., Granzer, Th., Scheck, M., \& Weber, M. 2000, \aaps, 142, 275
\bibitem[Torres(2007)]{Torres:07} Torres, G. 2007, \apjl, 671, L65
\bibitem[Torres et al.(2009)]{Torresetal:09} Torres, G., Andersen, J., \& Gim\'enez, A. 2009, \aapr, in press (arXiv:0908.2624v1)
\bibitem[Torres et al.(2000)]{Torres:00} Torres, G., Andersen, J., Nordstr\"om, B., \& Latham, D.\ W. 2000, \aj, 119, 1942
\bibitem[Torres \& Lacy(2009)]{Torres:09} Torres, G., \& Lacy, C.\ H.\ S. 2009, \aj, 137, 507
\bibitem[Torres et al.(2006)]{Torres:06} Torres G., Lacy C.\ H.\ S., Marschall L.\ A., Sheets H.\ A., Mader J.\ A. 2006, \apj, 640, 1018
\bibitem[Torres et al.(2002)]{Torres:02} Torres, G., Neuh\"auser, R., \& Guenther, E.\ W. 2002, \aj, 123, 1701
\bibitem[Torres \& Ribas(2002)]{Torres2002} Torres, G., \& Ribas, I. 2002, \apj, 567, 1140
\bibitem[Torres et al.(1997)]{Torres:97} Torres, G., Stefanik, R.\ P., Andersen, J., Nordstr\"om, B., Latham, D.\ W., \& Clausen. J.\ V. 1997, \aj, 114, 2764
\bibitem[Torres et al.(2008)]{Torres:08} Torres, G., Vaz, L.\ P.\ R., \& Lacy, C.\ H.\ S. 2008, \aj, 136, 2158
\bibitem[van Hamme \& Wilson(2003)]{vanHamme:03} van Hamme, W., \& Wilson, R.\ E. 2003, in GAIA Spectroscopy: Science and Technology, ASP Conf.\ Ser.\ 298, ed.\ U.\ Munari (San Francisco: ASP), p.\ 323
\bibitem[Wilson(1979)]{wilson1979} Wilson, R.\ E. 1979, \apj, 234, 1054
\bibitem[Wilson \& Devinney(1971)]{WD1971} Wilson, R.~E., \& Devinney, E.~J.\ 1971, \apj, 166, 605
\bibitem[Yi et al.(2001)]{Yi:01} Yi, S.\ K., Demarque, P., Kim, Y.-C., Lee, Y.-W., Ree, C.\ H., Lejeune, T., \& Barnes, S. 2001, \apjs, 136, 417
\bibitem[Zucker \& Mazeh(1994)]{Zucker:94} Zucker, S., \& Mazeh, T. 1994, \apj, 420, 806

\end{thebibliography}
\end{document}